\documentclass[preprint]{ptephy_v1}%%%%where ptephy_v1 is the template name

\preprintnumber{KIAS-P16067, KOBE-TH-16-06} %%% %%% Insert preprint number here

\usepackage{extarrows}
\usepackage{epsf}
\usepackage{color}
\usepackage{amsmath}
\usepackage{amssymb}
\usepackage{latexsym}
\usepackage{braket}
\usepackage{slashed}
\usepackage{float}
\usepackage{cases}
\usepackage{bm}
\usepackage{arydshln}
\usepackage{multirow}

\newcommand{\bp}{\begin{pmatrix}}
\newcommand{\ep}{\end{pmatrix}}
\newcommand{\bb}{\begin{bmatrix}}
\newcommand{\eb}{\end{bmatrix}}

\newcommand{\GkI}{I}
\newcommand{\GkII}{I\hspace{-.1em}I}
\newcommand{\GkIII}{I\hspace{-.1em}I\hspace{-.1em}I}

\newcommand{\varParallel}{\def\@varParallel{/\kern-.2em /}%
  \mathchoice%
  {\@varParallel}%
  {\textstyle\@varParallel}%
  {\scriptscriptstyle\@varParallel}%
  {\scriptscriptstyle\@varParallel}}

\definecolor{mygrn}{rgb}{0,0.6,0}

\newcommand{\cred}[1]{{\color{black}#1}} % : red -> black

\usepackage[normalem]{ulem}

\newcommand{\beq}{\begin{equation}}
\newcommand{\eeq}{\end{equation}}
\newcommand{\bea}{\begin{eqnarray}}
\newcommand{\eea}{\end{eqnarray}}

\newcommand{\fulltoday}{\number\day\space \ifcase\month\or
    January\or February\or March\or April\or May\or June\or
    July\or August\or September\or October\or November\or December\fi
    \space\number\year}

 \makeatletter
    
    \@addtoreset{equation}{section}
  \makeatother

\begin{document}

\title{Supersymmetry in \cred{the 6D Dirac action}}

%%%% To generate auto affiliation numbers please use \author{}\affil{} command

%\author{\name{\fname{Yukihiro} \surname{Fujimoto}}{1,\ast},
%\name{\fname{Kouhei} \surname{Hasegawa}}{2,\ast,\dag},
%\name{\fname{Kenji} \surname{Nishiwaki}}{3,\ast,\dag},

%\name{\fname{Makoto} \surname{Sakamoto}}{2,\ast,\dag}, and
%\name{\fname{Kentaro} \surname{Tatsumi}}{2,\ast,}\thanks{These authors contributed equally to this work}}
%%%\name{}{} Insert author name in first group and
%%% affiliation number in second group
%%% \fname for author firstname
%%% \surname for author surname
%%% \midname for author middle names
%\address{
%\affil{1}{${}^{1}$National Institute of Technology, Oita College, Oita 870-0152, Japan} \\
%\affil{2}{${}^{2}$Department of Physics, Kobe University, Kobe 657-8501, Japan} \\
%\affil{3}{${}^{3}$School of Physics, Korea Institute for Advanced Study~(KIAS), Seoul 02455, Republic of Korea}
%\email{134s110s@stu.kobe-u.ac.jp, kouhei@phys.sci.kobe-u.ac.jp, nishiken@kias.re.kr, dragon@kobe-u.ac.jp, 134s110s@stu.kobe-u.ac.jp}
%}
%%%% To generate auto affiliation numbers please use \author{}\affil{} command

\author{Yukihiro Fujimoto}
\affil{National Institute of Technology, Oita College, Oita 870-0152, Japan \email{y-fujimoto@oita-ct.ac.jp}}

\author{Kouhei Hasegawa}
\affil{Department of Physics, Kobe University, Kobe 657-8501, Japan \email{kouhei@phys.sci.kobe-u.ac.jp}}

\author{Kenji Nishiwaki}
\affil{School of Physics, Korea Institute for Advanced Study, Seoul 02455,
Republic of Korea \email{nishiken@kias.re.kr}}

\author{Makoto Sakamoto}
\affil{Department of Physics, Kobe University, Kobe 657-8501, Japan \email{dragon@kobe-u.ac.jp}}

\author{Kentaro Tatsumi}
\affil{Department of Physics, Kobe University, Kobe 657-8501, Japan \email{134s110s@stu.kobe-u.ac.jp}}

\begin{comment}
\author{Yukihiro Fujimoto}
\affil{National Institute of Technology, Oita College, Oita 870-0152, Japan}
%\email{134s110s@stu.kobe-u.ac.jp}

\author{Kouhei Hasegawa}
\affil{Department of Physics, Kobe University, Kobe 657-8501, Japan}
%\email{kouhei@phys.sci.kobe-u.ac.jp}

\author{Kenji Nishiwaki}
\affil{School of Physics, Korea Institute for Advanced Study~(KIAS), Seoul 02455,
Republic of Korea}
%\email{nishiken@kias.re.kr}

\author{Makoto Sakamoto}
\affil{Department of Physics, Kobe University, Kobe 657-8501, Japan}
%\email{dragon@kobe-u.ac.jp}

\author{Kentaro Tatsumi}
\affil{Department of Physics, Kobe University, Kobe 657-8501, Japan}
%\email{134s110s@stu.kobe-u.ac.jp}
\end{comment}

%%% To include the collaborator name... Please use the command "\collaborator"
%%% For example: \collaborator{ATLAS Collaboration}

\begin{abstract}%
We investigate a 6d Dirac fermion on a rectangle. It is found that the 4d spectrum is 
governed by $N=2$ supersymmetric quantum mechanics. Then we demonstrate that the 
supersymmetry is very useful \cred{for classifying all the} allowed boundary conditions and to expand 
the 6d Dirac field in \cred{Kaluza--Klein} modes. A striking feature of the model is that even 
though the 6d Dirac fermion has non-vanishing bulk mass, the 4d mass spectrum can contain
degenerate massless chiral fermions, which may provide a hint to solve \cred{the 
problem of the generation of} quarks and leptons. It is pointed out that \cred{zero-energy} solutions are 
not affected by the presence of the boundaries, while the boundary conditions 
work well for determining the \cred{positive-energy} solutions.
{We also provide a brief discussion on possible boundary conditions in \cred{the} general case, especially those on polygons.}
\end{abstract}

\subjectindex{B15, B33}

\maketitle

\section{Introduction
\label{sec:introduction}}

 The standard model has been completely established by the discovery of the Higgs boson 
 \cite{Aad:2012tfa, Chatrchyan:2012xdj}\cred{, and describes well the low-energy} physics below 
 the weak scale. Despite the great success of the standard model, it will be natural to 
 regard the standard model as a \cred{low-energy} effective theory of some more fundamental theories 
 defined at \cred{higher energy scales}. This is because the standard model leaves various problems
to be solved.

\cred{Promising} candidates beyond the standard model \cred{are} the models on \cred{higher-dimensional
space-times} with compact extra dimensions. \cred{These} models could solve the generation problem 
\cite{Libanov:2000uf, Frere:2000dc, Neronov:2001qv, Aguilar:2006sz, Gogberashvili:2007gg, 
Guo:2008ia, Kaplan:2011vz} and the fermion mass hierarchy one \cite{ArkaniHamed:1999dc, 
Dvali:2000ha, Gherghetta:2000qt, 1126-6708-2000-06-020, Kaplan:2001ga, Huber:2000ie, 
Kakizaki:2001ue}{,} and naturally explain the quark and lepton flavor structure 
\cite{Haba:2006dz, Abe:2008sx, Csaki:2008qq, Abe:2012fj} of the standard model. Many 
proposals have been made to explain the quark and lepton mass hierarchies and their 
flavor structures naturally from an extra-dimensional point of view.

Although extra-dimensional models will be expected to solve the generation problem, 
phenomenologically realistic models that \cred{solve} the problem are very limited. 
A possible mechanism \cred{for producing} degenerate massless chiral fermions is to put extra 
dimensions in a homogeneous magnetic field \cite{RandjbarDaemi:1982hi, Cremades:2004wa, 
Abe:2008fi, Abe:2008sx, Abe:2012fj, Fujimoto:2013xha, Abe:2013bca, Abe:2014noa, Abe:2015yva, 
Matsumoto:2016okl, Fujimoto:2016zjs}. Another mechanism is to put point interactions on 
an extra dimension \cite{Fujimoto:2012wv, Fujimoto:2013ki, Fujimoto:2014fka, Cai:2015jla}. 
It would be desirable to find new mechanism that \cred{solves} the above problems of the standard model and that can lead to phenomenologically realistic 
models with a simple setup.
%%%%%%%%% removed by sakamoto %%%%%%%%%%%%
%\textcolor{red}{, e.g. Higher-dimensional Dirac fermion with boundary conditions.}
%%%%%%%%%%%%%%%%%%%%%%%%%%%%%%%%%%%%%%%%%%

In the context of a five-dimensional (5d) gauge theory, it has been \cred{shown} that a 4d massless chiral fermion appears from a 5d Dirac {fermion} with a suitable boundary condition {(see e.g. \cite{Fujimoto:2011kf})}. 
Furthermore, {the} 5d Dirac mass term plays an important role \cred{in the} localization of {zero-mode} functions.
{Thereby, it can} become a source of the {observed} fermion mass hierarchy. 
\cred{Unfortunately, however,} in the case of 5d, only one 4d chiral fermion appears from a 5d Dirac field.
{On the other hand,} it would be expected that several 4d massless chiral fermions may emerge in \cred{the case of a} higher-dimensional Dirac fermion \cred{more than 5d that contains} more {degrees} of freedom than {those in 5d}.
Our {goal} is to solve the generation problem as well as other problems {in} the standard model from a higher-dimensional Dirac action point of view.

%%%%%%%%% New! added by sakamoto %%%%%%%%%%%%
In Ref.~{\cite{Fujimoto:2016llj}}, the 4d mass spectrum of a 6d Dirac fermion
\cred{was} investigated.
An interesting observation is that two 4d massless chiral fermions can appear,
even though the 6d Dirac action contains a non-zero bulk mass $M$.
The results strongly suggest that higher-dimensional Dirac fermions can 
provide more than two 4d massless chiral fermions and could solve the
generation problem.
Unfortunately, \cred{it is not} straightforward to extend the analysis
given in Ref.~{\cite{Fujimoto:2016llj}} to \cred{the} higher-dimensional Dirac action,
because the origin of \cred{the} degeneracy of the 4d mass spectrum (\cred{four} for \cred{the} massive modes\cred{,
and two} for \cred{the} massless modes) has been obscure\cred{, and it is especially} unclear
how to expand Dirac fields into \cred{Kaluza--Klein} modes for general 
\cred{higher dimensions}.

In this paper, we revisit the 6d Dirac fermion and reveal hidden structures
in the 4d mass spectrum from a symmetry point of view, in great detail.
We show that the 4d mass spectrum is governed by an $N=2$ quantum-mechanical
supersymmetry\cred{,} and the degeneracy of the 4d mass spectrum can be explained
by the supersymmetry (with an additional symmetry of the action).
This supersymmetric structure makes it clear why the 4d massless zero modes
become chiral.
This is because 4d massive modes always form supermultiplets and then become
Dirac fermions, but each massless zero mode does not form a supermultiplet
and hence has no chiral partner to form a Dirac fermion.
We further find that the supersymmetry is very powerful \cred{for analyzing the
Kaluza--Klein} mode expansions and \cred{determining} the class of allowed 
boundary conditions on extra dimensions.
We expect that our analysis can apply for general higher-dimensional
Dirac fermions and hence hope to answer the question \cred{of} whether or not Dirac
fermions with more than two extra dimensions can solve the \cred{generation
and fermion mass hierarchy problems}.
%%%%%%%%%%%%%%%%%%%%%%%%%%%%%%%%%%%%%%%%%%

%%%%%%%%% removed by sakamoto %%%%%%%%%%%%
%In this paper, \textcolor{red}{with considering the above-mentioned circumstances,} we investigate a 6d Dirac fermion with a bulk mass $M$. 
%In Ref.~{\cite{Fujimoto:2016llj}, the possible boundary conditions, 
%the mode functions and the mass spectrum are obtained in the same model.
%Then it has been shown that the degenerate chiral fermions appear in the 4d mass spectrum.
%Thus, it will be important to investigate the model in further details and to clarify a 
%hidden structure in the 4d mass spectrum, if exists. In fact, we reveal the supersymmetric 
%structure which governs the 4d mass spectrum and show that the supersymmetry is very useful 
%to analyze the Kaluza-Klein mode expansion.
%In particular, the difference between the massless chiral zero
%modes and the massive modes becomes clear in a supersymmetric point of view,
%because the massless (massive) modes turn out to correspond to
%solutions of the 1st order (2nd order) differential equation. Furthermore, we find that 
%the supersymmetric properties are powerful to determine the class of allowed boundary 
%conditions on extra dimensions.
%%%%%%%%%%%%%%%%%%%%%%%%%%%%%%%%%%%%%%%%%%

It is interesting to \cred{note} that the supersymmetric structure is a common feature in 
extra dimensions. This is because similar supersymmetric structures have been found in 
higher-dimensional gauge and gravity theories 
\cite{DeWolfe:1999cp, Miemiec:2000eq, Lim:2005rc, Lim:2007fy, Lim:2008hi, Ohya:2010wf, 
Nagasawa:2011mu, Sakamoto:2012ew} (see also \cite{Williams:2012au,Burgess:2012pc}). Thus,
it would be of great interest to understand \cred{the} role of the supersymmetry in extra dimensions
\cred{thoroughly}.

{This} paper is organized as follows. {We first give the setup of our model} in 
Section~\ref{sec:6d Dirac fermion on a rectangle} and then show, in 
Section~\ref{sec:Hidden N=2 supersymmetry}, that $N=2$ supersymmetric quantum 
mechanics is hidden in the 6d Dirac equation. 
In Section~\ref{sec:Classification of allowed boundary conditions}, we classify \cred{the} allowed 
boundary conditions with the help of the supersymmetry. 
In Sections~\ref{sec:Energy spectrum for type II BC} and \ref{sec:Energy spectrum 
for type III BC}, we explicitly construct \cred{positive-energy} eigenfunctions and point out 
a problem \cred{in determining zero-energy} solutions. The degeneracy of \cred{positive-energy} states 
are explained from symmetry transformations in Section~\ref{sec:Mapping between degenerate states}.
{In section~\ref{sec:general_BC}, we provide a brief discussion on possible boundary conditions in \cred{the} general case, especially those on polygons.}
Section~\ref{sec:Conclusions and discussions} is devoted to conclusions and discussions.

%%%%%%%%%%%%%%%%%%%%%%%%%%%%%%%%%%%%%
%%%%%%%% Section 2 %%%%%%%%%%%%%%%%%%
%%%%%%%%%%%%%%%%%%%%%%%%%%%%%%%%%%%%%
\section{\cred{Six-dimensional} Dirac fermion on a rectangle
\label{sec:6d Dirac fermion on a rectangle}}

Let us start with the 6d Dirac action
	%%%%%%%%%%%%%%%%%%%%%%%%%%%%%%%%%%%%%%%%%%%%%%%%%%%%%%%%%%%%%%%%%%%%%%%%%%
	\begin{align}
	S=\int d^{4}x\int^{L_{1}}_{0}dy_{1}\int^{L_{2}}_{0}dy_{2}\overline{\Psi}(x,y)\Bigl[i\Gamma^{A}\partial_{A}-M\Bigr]\Psi (x,y), \label{Dirac-action}
	\end{align}
	%%%%%%%%%%%%%%%%%%%%%%%%%%%%%%%%%%%%%%%%%%%%%%%%%%%%%%%%%%%%%%%%%%%%%%%%%%
where $\Psi (x,y)$ is an \cred{eight}-component Dirac spinor in \cred{six} dimensions and $M$ is \cred{the} bulk mass of the Dirac fermion. The 6d space-time is taken to be the direct product of the 4d Minkowski space-time and the 2d rectangle. The coordinates of the 4d Minkowski space-time and the 2d rectangle are denoted by $x^{\mu}$ ($\mu=0,1,2,3$) and $y_{j}$ ($j=1,2$), respectively. {The domain of the rectangle is set as} $0\leq y_{1}\leq L_{1}$ and $0\leq y_{2}\leq L_{2}$.

The Dirac action~(\ref{Dirac-action}) leads to the Dirac equation
	%%%%%%%%%%%%%%%%%%%%%%%%%%%%%%%%%%%%%%%%%%%%%%%%%%%%%%%%%%%%%%%%%%%%%%%%%%
	\begin{align}
	\Bigl[i\Gamma^{\mu}\partial_{\mu}+i\Gamma^{y_{1}}\partial_{y_{1}}+i\Gamma^{y_{2}}\partial_{y_{2}}-M\Bigr]\Psi(x,y)=0. \label{Dirac-equation}
	\end{align}
	%%%%%%%%%%%%%%%%%%%%%%%%%%%%%%%%%%%%%%%%%%%%%%%%%%%%%%%%%%%%%%%%%%%%%%%%%%
The 6d gamma matrices $\Gamma^{A}$ ($A=0,1,2,3,y_{1},y_{2}$) are required to satisfy
	%%%%%%%%%%%%%%%%%%%%%%%%%%%%%%%%%%%%%%%%%%%%%%%%%%%%%%%%%%%%%%%%%%%%%%%%%%
	\begin{align}
	&\{ \Gamma^{A},\Gamma^{B}\} =-2\eta^{AB}\, \mathrm{I}_{8},\qquad (A,B=0,1,2,3,y_{1},y_{2}),\nonumber\\
	&(\Gamma^{A})^{\dagger}=\left\{
									\begin{array}{ll}
									+\Gamma^{A}&A=0{,} \\
									-\Gamma^{A}&A\neq 0{,}
									\end{array}\right.
	\end{align}
	%%%%%%%%%%%%%%%%%%%%%%%%%%%%%%%%%%%%%%%%%%%%%%%%%%%%%%%%%%%%%%%%%%%%%%%%%%
with the 6d metric ${\rm diag}\,\eta^{AB}=(-1,1,1,1,1,1)$. Here, ${\rm I}_{n}$ denotes {the} $n\times n$ identity matrix. The Dirac conjugate $\overline{\Psi}$ is defined by $\overline{\Psi}=\Psi^{\dagger}\Gamma^{0}$, as usual.

In order to extract a quantum-mechanical supersymmetric structure from the Dirac equation~(\ref{Dirac-equation}), {it may be necessary to drive the equation without including the gamma matrices $\Gamma^{y_{1}}$ and $\Gamma^{y_{2}}$}. {{For this purpose,} it turns out \cred{to be} convenient to introduce the matrices $\Gamma^{5}$ and $\Gamma^{y}$ such as}
	%%%%%%%%%%%%%%%%%%%%%%%%%%%%%%%%%%%%%%%%%%%%%%%%%%%%%%%%%%%%%%%%%%%%%%%%%%
	\begin{align}
	\Gamma^{5} \ & {\equiv} \ i\Gamma^{0}\Gamma^{1}\Gamma^{2}\Gamma^{3},\\
	\Gamma^{y} \ & {\equiv}\  i\Gamma^{y_{1}}\Gamma^{y_{2}}{,}
	\end{align}
	%%%%%%%%%%%%%%%%%%%%%%%%%%%%%%%%%%%%%%%%%%%%%%%%%%%%%%%%%%%%%%%%%%%%%%%%%%
where $\Gamma^{y}$ is an analogue of $\gamma^{5}$ in the extra dimensions.

Since $\Gamma^{y}$ commutes with $\Gamma^{5}$, we can introduce simultaneous eigenstates of $\Gamma^{5}$ and $\Gamma^{y}$ defined by 
	%%%%%%%%%%%%%%%%%%%%%%%%%%%%%%%%%%%%%%%%%%%%%%%%%%%%%%%%%%%%%%%%%%%%%%%%%%
	\begin{align}
	\Gamma^{5}\Psi_{R\pm}=+\Psi_{R\pm},\qquad \Gamma^{5}\Psi_{L\pm}=-\Psi_{L\pm},\\
	\Gamma^{y}\Psi_{R\pm}=\pm\Psi_{R\pm},\qquad \Gamma^{y}\Psi_{L\pm}=\pm\Psi_{L\pm}.
	\end{align}
	%%%%%%%%%%%%%%%%%%%%%%%%%%%%%%%%%%%%%%%%%%%%%%%%%%%%%%%%%%%%%%%%%%%%%%%%%%
By use of the projection matrices, $\Psi_{R\pm}$ and $\Psi_{L\pm}$ can be constructed from $\Psi$ as
	%%%%%%%%%%%%%%%%%%%%%%%%%%%%%%%%%%%%%%%%%%%%%%%%%%%%%%%%%%%%%%%%%%%%%%%%%%
	\begin{align}
	{\Psi_{R\pm}\equiv {\cal P}_{R}{\cal P}_{\pm}\Psi},\qquad \Psi_{L\pm}\equiv {\cal P}_{L}{\cal P}_{\pm}\Psi,
	\end{align}
	%%%%%%%%%%%%%%%%%%%%%%%%%%%%%%%%%%%%%%%%%%%%%%%%%%%%%%%%%%%%%%%%%%%%%%%%%%
where
	%%%%%%%%%%%%%%%%%%%%%%%%%%%%%%%%%%%%%%%%%%%%%%%%%%%%%%%%%%%%%%%%%%%%%%%%%%
	\begin{align}
	{\cal P}_{R}&=\frac{1}{2}({\rm I}_{8}+\Gamma^{5}),\qquad {\cal P}_{L}=\frac{1}{2}({\rm I}_{8}-\Gamma^{5}),\\
	{\cal P}_{\pm}&=\frac{1}{2}({\rm I}_{8}\pm\Gamma^{y}).
	\end{align}
	%%%%%%%%%%%%%%%%%%%%%%%%%%%%%%%%%%%%%%%%%%%%%%%%%%%%%%%%%%%%%%%%%%%%%%%%%%

In terms of the eigenstates of $\Gamma^{5}$ and $\Gamma^{y}$, the Dirac equation (\ref{Dirac-equation}) can be decomposed as
	%%%%%%%%%%%%%%%%%%%%%%%%%%%%%%%%%%%%%%%%%%%%%%%%%%%%%%%%%%%%%%%%%%%%%%%%%%
	\begin{align}
	i\Gamma^{\mu}\partial_{\mu}\Psi_{R+}=-i\Gamma^{y_{1}}(\partial_{y_{1}}-i\partial_{y_{2}})\Psi_{L-}+M\Psi_{L+},\nonumber\\
	i\Gamma^{\mu}\partial_{\mu}\Psi_{R-}=-i\Gamma^{y_{1}}(\partial_{y_{1}}+i\partial_{y_{2}})\Psi_{L+}+M\Psi_{L-},\nonumber\\
	i\Gamma^{\mu}\partial_{\mu}\Psi_{L+}=-i\Gamma^{y_{1}}(\partial_{y_{1}}-i\partial_{y_{2}})\Psi_{R-}+M\Psi_{R+},\nonumber\\
	i\Gamma^{\mu}\partial_{\mu}\Psi_{L-}=-i\Gamma^{y_{1}}(\partial_{y_{1}}+i\partial_{y_{2}})\Psi_{R+}+M\Psi_{R-}{.}
	\label{decomposed-Dirad-equations}
	\end{align}
	%%%%%%%%%%%%%%%%%%%%%%%%%%%%%%%%%%%%%%%%%%%%%%%%%%%%%%%%%%%%%%%%%%%%%%%%%%
Furthermore, in order to remove $i\Gamma^{y_{1}}$ from the above equations, we may redefine the fields $\Psi_{R\pm}$ and $\Psi_{L\pm}$ as
	%%%%%%%%%%%%%%%%%%%%%%%%%%%%%%%%%%%%%%%%%%%%%%%%%%%%%%%%%%%%%%%%%%%%%%%%%%
	\begin{align}
	\Psi_{R1}\equiv \Psi_{R+},\qquad \Psi_{R2}\equiv i\Gamma^{y_{1}}\Psi_{R-},\nonumber\\
	\Psi_{L1}\equiv \Psi_{L+},\qquad \Psi_{L2}\equiv i\Gamma^{y_{1}}\Psi_{L-}.
	\label{eq:new_basis_fermion}
	\end{align}
	%%%%%%%%%%%%%%%%%%%%%%%%%%%%%%%%%%%%%%%%%%%%%%%%%%%%%%%%%%%%%%%%%%%%%%%%%%
{Then, we have succeeded in eliminating the gamma matrix} $\Gamma^{y_{1}}$ from (\ref{decomposed-Dirad-equations}) and in rewriting (\ref{decomposed-Dirad-equations}) into the form
	%%%%%%%%%%%%%%%%%%%%%%%%%%%%%%%%%%%%%%%%%%%%%%%%%%%%%%%%%%%%%%%%%%%%%%%%%%
	\begin{align}
	i\Gamma^{\mu}\partial_{\mu}\left(\begin{array}{c}
									\Psi_{R1}(x,y)\\
									\Psi_{R2}(x,y)\\
									\Psi_{L1}(x,y)\\
									\Psi_{L2}(x,y)
									\end{array}\right)= (Q\otimes {\rm I}_{2})\left(\begin{array}{c}
									\Psi_{R1}(x,y)\\
									\Psi_{R2}(x,y)\\
									\Psi_{L1}(x,y)\\
									\Psi_{L2}(x,y)
									\end{array}\right){,}
	\label{New-Dirac-equations}
	\end{align}
	%%%%%%%%%%%%%%%%%%%%%%%%%%%%%%%%%%%%%%%%%%%%%%%%%%%%%%%%%%%%%%%%%%%%%%%%%%
where ${\rm I}_{2}$ acts on two-dimensional spinors $\Psi_{R1}$, $\Psi_{R2}$, $\Psi_{L1}$, $\Psi_{L2}$, and the $4\times 4$ matrix $Q$ is defined by 
	%%%%%%%%%%%%%%%%%%%%%%%%%%%%%%%%%%%%%%%%%%%%%%%%%%%%%%%%%%%%%%%%%%%%%%%%%%
	\begin{align}
	Q\equiv \left(\begin{array}{cccc}
				0&0&M&-(\partial_{y_{1}}-i\partial_{y_{2}})\\
				0&0&\partial_{y_{1}}+i\partial_{y_{2}}&-M\\
				M&{-(\partial_{y_{1}}-i\partial_{y_{2}})}&0&0\\
				\partial_{y_{1}}+i\partial_{y_{2}}&-M&0&0
				\end{array}\right).\label{supercharge}
	\end{align}
	%%%%%%%%%%%%%%%%%%%%%%%%%%%%%%%%%%%%%%%%%%%%%%%%%%%%%%%%%%%%%%%%%%%%%%%%%%
It should be emphasized that $Q$ does not act on spinor indices but on the ``flavor'' space displayed in (\ref{New-Dirac-equations}) and satisfies the relation
	%%%%%%%%%%%%%%%%%%%%%%%%%%%%%%%%%%%%%%%%%%%%%%%%%%%%%%%%%%%%%%%%%%%%%%%%%%
	\begin{align}
	Q^{2}=\Bigl[-\partial_{y_{1}}^{2}-\partial_{y_{2}}^{2}+M^{2}\Bigr]{\rm I}_{4}.
	\end{align}
	%%%%%%%%%%%%%%%%%%%%%%%%%%%%%%%%%%%%%%%%%%%%%%%%%%%%%%%%%%%%%%%%%%%%%%%%%%
Thus, the differential operator $Q^{2}$ turns out to correspond to a Laplacian on the extra dimensions.

In the following sections, we will show that $Q$ can be regarded as a supercharge of $N=2$ supersymmetric quantum mechanics and that the 4d mass spectrum of the 6d Dirac fermion system is governed by the supersymmetry.

%%%%%%%%%%%%%%%%%%%%%%%%%%%%%%%%%%%%%
%%%%%%%% Section 3 %%%%%%%%%%%%%%%%%%
%%%%%%%%%%%%%%%%%%%%%%%%%%%%%%%%%%%%%
\section{Hidden $N=2$ supersymmetry
\label{sec:Hidden N=2 supersymmetry}}

Since we would like to regard $Q$ as a supercharge in supersymmetric quantum mechanics, we may introduce a Hamiltonian $H$ by
	%%%%%%%%%%%%%%%%%%%%%%%%%%%%%%%%%%%%%%%%%%%%%%%%%%%%%%%%%%%%%%%%%%%%%%%%%%
	\begin{align}
	H=Q^{2}.
	\end{align}
	%%%%%%%%%%%%%%%%%%%%%%%%%%%%%%%%%%%%%%%%%%%%%%%%%%%%%%%%%%%%%%%%%%%%%%%%%%
In order for the system to be supersymmetric, we further need to introduce the ``fermion'' number operator $F$ which should satisfy the relation \cite{Witten:1981nf}
	%%%%%%%%%%%%%%%%%%%%%%%%%%%%%%%%%%%%%%%%%%%%%%%%%%%%%%%%%%%%%%%%%%%%%%%%%%
	\begin{align}
	(-1)^{F}Q&=-Q(-1)^{F},\nonumber\\
	\Bigl[(-1)^{F}\Bigr]^{2}&={\rm I}_{4}.\label{SUSY-algebra}
	\end{align}
	%%%%%%%%%%%%%%%%%%%%%%%%%%%%%%%%%%%%%%%%%%%%%%%%%%%%%%%%%%%%%%%%%%%%%%%%%%
Then, the operator $H$, $Q$ and $(-1)^{F}$ are assumed to act on \cred{four}-component wavefunctions
	%%%%%%%%%%%%%%%%%%%%%%%%%%%%%%%%%%%%%%%%%%%%%%%%%%%%%%%%%%%%%%%%%%%%%%%%%%
	\begin{align}
	\Phi(y)=\left(\begin{array}{c}
					f_{1}(y)\\
					f_{2}(y)\\
					g_{1}(y)\\
					g_{2}(y)
					\end{array}\right)\label{wavefunctions}
	\end{align}
	%%%%%%%%%%%%%%%%%%%%%%%%%%%%%%%%%%%%%%%%%%%%%%%%%%%%%%%%%%%%%%%%%%%%%%%%%%
\cred{that} depend only on $y_{1}$ and $y_{2}$.

The operator $(-1)^{F}$ obeying the relations (\ref{SUSY-algebra}) is found to be of the form
	%%%%%%%%%%%%%%%%%%%%%%%%%%%%%%%%%%%%%%%%%%%%%%%%%%%%%%%%%%%%%%%%%%%%%%%%%%
	\begin{align}
	(-1)^{F}=\left( \begin{array}{cccc}
					1&0&0&0\\
					0&1&0&0\\
					0&0&-1&0\\
					0&0&0&-1
					\end{array}\right).
	\end{align}
	%%%%%%%%%%%%%%%%%%%%%%%%%%%%%%%%%%%%%%%%%%%%%%%%%%%%%%%%%%%%%%%%%%%%%%%%%%
In the context of supersymmetry, we might call $(-1)^{F}=+1$ ($-1$) eigenstates ``bosonic'' (``fermionic'') states, though they do not literally mean bosonic or fermionic states in our model. It is worth noting that the eigenstates of $(-1)^{F}=+1$ ($-1$) rather correspond to those of $\Gamma^{5}=+1$ ($-1$) from (\ref{New-Dirac-equations}), so that $(-1)^{F}$ may be regarded as a counterpart of the 4d chiral operator.

The Hamiltonian system equipped with $Q$ and $(-1)^{F}$ is called an $N=2$ supersymmetric quantum mechanics\hspace{-0.3em}
	%%%%%%%%%%%%%%%%	footnote	%%%%%%%%%%%%%%%%%%
	~\footnote{If we want to have two supercharges, we may introduce them by $Q_{1}\equiv Q$ and $Q_{2}\equiv -iQ(-1)^{F}$. Then, we can show that they form the $N=2$ supersymmetry algebra, i.e. $\{Q_{i},Q_{j}\}=2H\delta_{ij}$ ($i,j=1,2$).}
	%%%%%%%%%%%%%%%%%%%%%%%%%%%%%%%%%%%%%%%%%%%%%%%%%%
or a Witten model \cite{Witten:1981nf, Cooper:1994eh, Cooper:2001zd} {\it if} $Q$ and $(-1)^{F}$ are \cred{Hermitian}, i.e.
	%%%%%%%%%%%%%%%%%%%%%%%%%%%%%%%%%%%%%%%%%%%%%%%%%%%%%%%%%%%%%%%%%%%%%%%%%%
	\begin{align}
	&Q^{\dagger}=Q,\label{supercharge-hermiticity}\\
	&\Bigl[(-1)^{F}\Bigr]^{\dagger}=(-1)^{F}.
	\end{align}
	%%%%%%%%%%%%%%%%%%%%%%%%%%%%%%%%%%%%%%%%%%%%%%%%%%%%%%%%%%%%%%%%%%%%%%%%%%

It should be emphasized that the above \cred{Hermiticity} property of $Q$ is {\it not} trivial because the extra dimensions have boundaries. In fact, we will {see} in the next section that the \cred{Hermiticity} requirement (\ref{supercharge-hermiticity}) and the compatibility condition with $(-1)^{F}$ severely restrict \cred{the} allowed boundary conditions for the wavefunctions (\ref{wavefunctions}) at the boundaries of the rectangle.

%%%%%%%%%%%%%%%%%%%%%%%%%%%%%%%%%%%%%
%%%%%%%% Section 4 %%%%%%%%%%%%%%%%%%
%%%%%%%%%%%%%%%%%%%%%%%%%%%%%%%%%%%%%
\section{Classification of allowed boundary conditions
\label{sec:Classification of allowed boundary conditions}}

%%%%%%%%%%%%%%%		Subsection~4.1		%%%%%%%%%%%%%%%%%%%
\subsection{Requirement of \cred{Hermiticity} for $Q$}
\label{sec:Requirement of hermiticity for $Q$}

Since we have taken the extra dimensions to be a rectangle, the requirement \cred{for Hermiticity for} the supercharge $Q$ is not trivial. In fact, we will see below that the \cred{Hermiticity} requirement severely restricts the class of allowed boundary conditions for $\Phi (y)$ at $y_{1}=0, L_{1}$ and $y_{2}=0,L_{2}$.

To be more precise, we require that the supercharge $Q$ is \cred{Hermitian} under the inner product
	%%%%%%%%%%%%%%%%%%%%%%%%%%%%%%%%%%%%%%%%%%%%%%%%%%%%%%%%%%%%%%%%%%%%%%%%%%
	\begin{align}
	\langle \Phi' ,\Phi\rangle &\equiv \int^{L_{1}}_{0}dy_{1}\int^{L_{2}}_{0}dy_{2}
	{\Bigl(\Phi'(y)\Bigr)^{\dagger}} \Phi(y)\nonumber\\
	&=\int^{L_{1}}_{0}dy_{1}\int^{L_{2}}_{0}dy_{2}\Bigl\{\Bigl(f'_{1}(y)\Bigr)^{\ast}f_{1}(y)+\Bigl(f'_{2}(y)\Bigr)^{\ast}f_{2}(y)+\Bigl(g'_{1}(y)\Bigr)^{\ast}g_{1}(y)+\Bigl(g'_{2}(y)\Bigr)^{\ast}g_{2}(y)\Bigr\},
	\end{align}
	%%%%%%%%%%%%%%%%%%%%%%%%%%%%%%%%%%%%%%%%%%%%%%%%%%%%%%%%%%%%%%%%%%%%%%%%%%
where
	%%%%%%%%%%%%%%%%%%%%%%%%%%%%%%%%%%%%%%%%%%%%%%%%%%%%%%%%%%%%%%%%%%%%%%%%%%
	\begin{align}
		\Phi(y)=\left(\begin{array}{c}
					f_{1}(y)\\
					f_{2}(y)\\
					g_{1}(y)\\
					g_{2}(y)
					\end{array}\right),\qquad 	\Phi'(y)=\left(\begin{array}{c}
					f'_{1}(y)\\
					f'_{2}(y)\\
					g'_{1}(y)\\
					g'_{2}(y)
					\end{array}\right).
	\end{align}
	%%%%%%%%%%%%%%%%%%%%%%%%%%%%%%%%%%%%%%%%%%%%%%%%%%%%%%%%%%%%%%%%%%%%%%%%%%
Then, in order for $Q$ to be \cred{Hermitian}, $Q$ has to satisfy
	%%%%%%%%%%%%%%%%%%%%%%%%%%%%%%%%%%%%%%%%%%%%%%%%%%%%%%%%%%%%%%%%%%%%%%%%%%
	\begin{align}
	\langle Q\Phi', \Phi\rangle=\langle \Phi', Q\Phi\rangle \label{Innerproduct-Supercharge-Hermiticity}
	\end{align}
	%%%%%%%%%%%%%%%%%%%%%%%%%%%%%%%%%%%%%%%%%%%%%%%%%%%%%%%%%%%%%%%%%%%%%%%%%%
for arbitrary \cred{four}-component wavefunctions $\Phi(y)$ and $\Phi'(y)$ with appropriate boundary conditions.

To make our analysis tractable, we assume that the probability current in the directions of the extra dimensions \cred{terminates} at each point of the boundaries of the rectangle. Then, (\ref{Innerproduct-Supercharge-Hermiticity}) turns out to reduce to the conditions
	%%%%%%%%%%%%%%%%%%%%%%%%%%%%%%%%%%%%%%%%%%%%%%%%%%%%%%%%%%%%%%%%%%%%%%%%%%
	\begin{align}
	\Bigl(f'_{1}(y)\Bigr)^{\ast}g_{2}(y)-\Bigl(f'_{2}(y)\Bigr)^{\ast}g_{1}(y)+\Bigl(g'_{1}(y)\Bigr)^{\ast}f_{2}(y)-\Bigl(g'_{2}(y)\Bigr)^{\ast}f_{1}(y)=0\qquad \text{at}\ \ y_{1}=0,L_{1},\label{y1-direction-probability-current-BC}\\
	\Bigl(f'_{1}(y)\Bigr)^{\ast}g_{2}(y)+\Bigl(f'_{2}(y)\Bigr)^{\ast}g_{1}(y)+\Bigl(g'_{1}(y)\Bigr)^{\ast}f_{2}(y)+\Bigl(g'_{2}(y)\Bigr)^{\ast}f_{1}(y)=0\qquad \text{at}\ \ y_{2}=0,L_{2},\label{y2-direction-probability-current-BC}
	\end{align}
	%%%%%%%%%%%%%%%%%%%%%%%%%%%%%%%%%%%%%%%%%%%%%%%%%%%%%%%%%%%%%%%%%%%%%%%%%%

%%%%%%%%%%%%%%%		Subsection~4.2		%%%%%%%%%%%%%%%%%%%
\subsection{Allowed boundary conditions in the $y_{1}$-direction}
\label{sec:Allowed boundary conditions in the y_{1}-direction}

Let us first investigate \cred{condition} (\ref{y1-direction-probability-current-BC}). To solve \cred{condition} (\ref{y1-direction-probability-current-BC}), we first restrict our considerations to the case of $\Phi'(y)=\Phi(y)$, i.e. $f'_{j}(y)=f_{j}(y)$ and $g'_{j}(y)=g_{j}(y)$ ($j=1,2$). This restriction would give a necessary condition for (\ref{y1-direction-probability-current-BC}). We will, however, verify that \cred{the} derived boundary conditions are sufficient as well as necessary.

For $f'_{j}(y)=f_{j}(y)$ and $g'_{j}(y)=g_{j}(y)$ ($j=1,2$), \cred{condition} (\ref{y1-direction-probability-current-BC}) can be written in the form
	%%%%%%%%%%%%%%%%%%%%%%%%%%%%%%%%%%%%%%%%%%%%%%%%%%%%%%%%%%%%%%%%%%%%%%%%%%
	\begin{align}
	\rho_{1}(y)^{\dagger}\lambda_{1}(y)+\lambda_{1}(y)^{\dagger}\rho_{1}(y)=0\qquad \text{at}\ \ y_{1}=0,L_{1},\label{rho-lambda-boundaries}
	\end{align}
	%%%%%%%%%%%%%%%%%%%%%%%%%%%%%%%%%%%%%%%%%%%%%%%%%%%%%%%%%%%%%%%%%%%%%%%%%%
where $\rho_{1}(y)$ and $\lambda_{1}(y)$ are \cred{two}-component vectors defined by
	%%%%%%%%%%%%%%%%%%%%%%%%%%%%%%%%%%%%%%%%%%%%%%%%%%%%%%%%%%%%%%%%%%%%%%%%%%
	\begin{align}
	\rho_{1}(y)=\left(\begin{array}{c}
						f_{1}(y)\\
						f_{2}(y)
						\end{array}\right),\qquad \lambda_{1}(y) = \cred{i} \left(\begin{array}{c}
						-g_{2}(y)\\
						g_{1}(y)
						\end{array}\right).
	\end{align}
	%%%%%%%%%%%%%%%%%%%%%%%%%%%%%%%%%%%%%%%%%%%%%%%%%%%%%%%%%%%%%%%%%%%%%%%%%%
A crucial observation is that the condition (\ref{rho-lambda-boundaries}) can be rewritten as
	%%%%%%%%%%%%%%%%%%%%%%%%%%%%%%%%%%%%%%%%%%%%%%%%%%%%%%%%%%%%%%%%%%%%%%%%%%
	\begin{align}
	|\rho_{1}(y)+L_{0}\lambda_{1}(y)|^{2}=|\rho_{1}(y)-L_{0}\lambda_{1}(y)|^{2}\qquad \text{at}\ \ y_{1}=0,L_{1},\label{ablusolute-form-rho-lambda-boundaries}
	\end{align}
	%%%%%%%%%%%%%%%%%%%%%%%%%%%%%%%%%%%%%%%%%%%%%%%%%%%%%%%%%%%%%%%%%%%%%%%%%%
where $L_{0}$ is a non-zero real constant whose value is irrelevant unless $L_{0}$ is non-vanishing. {General solutions to (\ref{ablusolute-form-rho-lambda-boundaries}) are easily found in the form}
	%%%%%%%%%%%%%%%%%%%%%%%%%%%%%%%%%%%%%%%%%%%%%%%%%%%%%%%%%%%%%%%%%%%%%%%%%%
	\begin{align}
	\rho_{1}(y)+L_{0}\lambda_{1}(y)=U_{1} \Bigl(\rho_{1}(y)-L_{0}\lambda_{1}(y)\Bigr)\qquad \text{at}\ \ y_{1}=0,L_{1},\nonumber
	\end{align}
	%%%%%%%%%%%%%%%%%%%%%%%%%%%%%%%%%%%%%%%%%%%%%%%%%%%%%%%%%%%%%%%%%%%%%%%%%%
or equivalently
	%%%%%%%%%%%%%%%%%%%%%%%%%%%%%%%%%%%%%%%%%%%%%%%%%%%%%%%%%%%%%%%%%%%%%%%%%%
	\begin{align}
	\Bigl({\rm I}_{2}-U_{1}\Bigr)\rho_{1}(y)=-L_{0}\Bigl({\rm I}_{2}+U_{1}\Bigr)\lambda_{1}(y)\qquad \text{at}\ \ y_{1}=0,L_{1}, \label{y1-direction-generalBC}
	\end{align}
	%%%%%%%%%%%%%%%%%%%%%%%%%%%%%%%%%%%%%%%%%%%%%%%%%%%%%%%%%%%%%%%%%%%%%%%%%%
where $U_{1}$ is an arbitrary $2\times 2$ unitary matrix.

We have required the \cred{Hermiticity} of the supercharge $Q$ in order for the system to be supersymmetric. The \cred{Hermiticity} of $Q$ is{, however, not} enough to preserve {the} supersymmetry. We should further require that the boundary conditions are compatible with the fermion number operator $(-1)^{F}$.

Since $(-1)^{F}$ commutes with the Hamiltonian $H$, $(-1)^{F}$ can be regarded as a conserved charge. Hence, the eigenvalues of $(-1)^{F}$ should be conserved, otherwise the supersymmetric structure would be destroyed. Since $\rho_{1}(y)=\Bigl(f_{1}(y),f_{2}(y)\Bigr)^{\rm T}$ and $\lambda_{1}(y)=\Bigl(-g_{2}(y),g_{1}(y)\Bigr)^{\rm T}$ correspond to $(-1)^{F}=+1$ and $-1$, respectively, $\rho_{1}(y)$ should not be related to $\lambda_{1}(y)$ at the boundaries in order for eigenvalues of $(-1)^{F}$ to be conserved.\footnote{\cred{It is worth noting} that this requirement will correspond to that of the 4d Lorentz invariance in the original 6d action, as discussed in Ref.~{\cite{Fujimoto:2016llj}}.}
Therefore, the condition (\ref{y1-direction-generalBC}) has to reduce to
	%%%%%%%%%%%%%%%%%%%%%%%%%%%%%%%%%%%%%%%%%%%%%%%%%%%%%%%%%%%%%%%%%%%%%%%%%%
	\begin{align}
	&\Bigl({\rm I}_{2}-U_{1}\Bigr)\rho_{1}(y)=0,\qquad \text{at}\ \ y_{1}=0,L_{1},\label{y1-direction-rho-generalBC}\\
	&{\Bigl({\rm I}_{2} + U_{1}\Bigr)\lambda_{1}(y)=0},\qquad \text{at}\ \ y_{1}=0,L_{1}.\label{y1-direction-lambda-generalBC}
	\end{align}
	%%%%%%%%%%%%%%%%%%%%%%%%%%%%%%%%%%%%%%%%%%%%%%%%%%%%%%%%%%%%%%%%%%%%%%%%%%
In other words, only a class of $U_{1}$ that (\ref{y1-direction-generalBC}) \cred{reduces} to (\ref{y1-direction-rho-generalBC}) and (\ref{y1-direction-lambda-generalBC}) is permitted.

It is not difficult to show that the condition (\ref{y1-direction-generalBC}) can reduce to (\ref{y1-direction-rho-generalBC}) and (\ref{y1-direction-lambda-generalBC}) only if the eigenvalues of $U_{1}$ are equal to $+1$ or $-1$. This implies that the diagonalized form of $U_{1}$ can be categorized into \cred{three} types:\footnote{One might add the case of $U_{1}^{\rm diag}=\left(\begin{array}{cc}
																-1&0\\
																0&1
																\end{array}\right)$ to the list, but it turns out that this case leads to the same results as those of the type I\!I\!I.}
	%%%%%%%%%%%%%%%%%%%%%%%%%%%%%%%%%%%%%%%%%%%%%%%%%%%%%%%%%%%%%%%%%%%%%%%%%%
	\begin{description}
	\item[\r{1})]\ Type I
		%%%%%%%%%%%%%%%%%%%%%%%%%%%%%%%%%%%%%%%%%%%%%%%%%%%%%%%%%%%%%%%%%%%%%%%%%%
		\begin{align}
		U_{1}^{\rm diag}=\left(\begin{array}{cc}
								1&0\\
								0&1
								\end{array}\right){,}
		\label{typeI-BC}
		\end{align}
		%%%%%%%%%%%%%%%%%%%%%%%%%%%%%%%%%%%%%%%%%%%%%%%%%%%%%%%%%%%%%%%%%%%%%%%%%%
	\item[\r{2})]\ Type $\text{\GkII}$
		%%%%%%%%%%%%%%%%%%%%%%%%%%%%%%%%%%%%%%%%%%%%%%%%%%%%%%%%%%%%%%%%%%%%%%%%%%
		\begin{align}
		U_{1}^{\rm diag}=\left(\begin{array}{cc}
								-1&0\\
								0&-1
								\end{array}\right){,}
		\label{typeII-BC}
		\end{align}
		%%%%%%%%%%%%%%%%%%%%%%%%%%%%%%%%%%%%%%%%%%%%%%%%%%%%%%%%%%%%%%%%%%%%%%%%%%
	\item[\r{3})]\ Type $\text{\GkIII}$
		%%%%%%%%%%%%%%%%%%%%%%%%%%%%%%%%%%%%%%%%%%%%%%%%%%%%%%%%%%%%%%%%%%%%%%%%%%
		\begin{align}
		U_{1}^{\rm diag}=\left(\begin{array}{cc}
								1&0\\
								0&-1
								\end{array}\right){.}
		\label{typeIII-BC}
		\end{align}
		%%%%%%%%%%%%%%%%%%%%%%%%%%%%%%%%%%%%%%%%%%%%%%%%%%%%%%%%%%%%%%%%%%%%%%%%%%
	\end{description}
	%%%%%%%%%%%%%%%%%%%%%%%%%%%%%%%%%%%%%%%%%%%%%%%%%%%%%%%%%%%%%%%%%%%%%%%%%%
In the following, we will derive a general form of $U_{1}$ associated with each of (\ref{typeI-BC}), (\ref{typeII-BC}) and (\ref{typeIII-BC}).

	%%%%%%%%%%%%%%%%%%%%%%%%%%%%%%%%%%%%%%%%%%%%%%%%%%%%%%%%%%%%%%%%%%%%%%%%%%
	\begin{description}
	\item[\r{1})]\ \underline{Type I boundary condition}\\[0.3cm]
		The unitary matrix $U_{1}$ can be diagonalized by a unitary matrix $V$ such that
			%%%%%%%%%%%%%%%%%%%%%%%%%%%%%%%%%%%%%%%%%%%%%%%%%%%%%%%%%%%%%%%%%%%%%%%%%%
			\begin{align}
			VU_{1}V^{-1}=U^{\rm diag}_{1}.\label{typeI-BC-diagonalization}
			\end{align}
			%%%%%%%%%%%%%%%%%%%%%%%%%%%%%%%%%%%%%%%%%%%%%%%%%%%%%%%%%%%%%%%%%%%%%%%%%%
		Since $U^{\rm diag}_{1}$ is the identity matrix for \cred{the} {Type I case} of (\ref{typeI-BC}), (\ref{typeI-BC-diagonalization}) implies that $U_{1}$ is also identity matrix, i.e.
			%%%%%%%%%%%%%%%%%%%%%%%%%%%%%%%%%%%%%%%%%%%%%%%%%%%%%%%%%%%%%%%%%%%%%%%%%%
			\begin{align}
			U_{1}={\rm I}_{2}.
			\end{align}
			%%%%%%%%%%%%%%%%%%%%%%%%%%%%%%%%%%%%%%%%%%%%%%%%%%%%%%%%%%%%%%%%%%%%%%%%%%
		Then, the condition (\ref{y1-direction-rho-generalBC}) is trivially satisfied, and (\ref{y1-direction-lambda-generalBC}) reduces to
			%%%%%%%%%%%%%%%%%%%%%%%%%%%%%%%%%%%%%%%%%%%%%%%%%%%%%%%%%%%%%%%%%%%%%%%%%%
			\begin{align}
			g_{1}(y)=g_{2}(y)=0\qquad \text{at}\ \ y_{1}=0,L_{1}.\label{typeI-BC-g}
			\end{align}
			%%%%%%%%%%%%%%%%%%%%%%%%%%%%%%%%%%%%%%%%%%%%%%%%%%%%%%%%%%%%%%%%%%%%%%%%%%
		It will be convenient to rewrite the boundary condition (\ref{typeI-BC-g}), in terms of the original \cred{four}-component wavefunction $\Phi(y)$ \cred{as}
			%%%%%%%%%%%%%%%%%%%%%%%%%%%%%%%%%%%%%%%%%%%%%%%%%%%%%%%%%%%%%%%%%%%%%%%%%%
			\begin{align}
			{\cal P}_{(-1)^{F}=-1}\Phi(y)=0\qquad \text{at}\ \ y_{1}=0,L_{1},
			\end{align}
			%%%%%%%%%%%%%%%%%%%%%%%%%%%%%%%%%%%%%%%%%%%%%%%%%%%%%%%%%%%%%%%%%%%%%%%%%%	
	for \cred{the} {Type I} boundary condition. Here, ${\cal P}_{(-1)^{F}=-1}$ denotes the projection matrix defined by
			%%%%%%%%%%%%%%%%%%%%%%%%%%%%%%%%%%%%%%%%%%%%%%%%%%%%%%%%%%%%%%%%%%%%%%%%%%
			\begin{align}
			{\cal P}_{(-1)^{F}=\pm 1}\equiv \frac{1}{2}\Bigl({\rm I}_{4}\pm (-1)^{F}\Bigr).
			\end{align}
			%%%%%%%%%%%%%%%%%%%%%%%%%%%%%%%%%%%%%%%%%%%%%%%%%%%%%%%%%%%%%%%%%%%%%%%%%%
					
	\item[\r{2})]\ \underline{Type {\GkII} boundary condition}\\[0.3cm]
		Since $U^{\rm diag}_{1}$ given in (\ref{typeII-BC}) is proportional to the identity matrix, the unitary matrix $U_{1}$ is given by 
			%%%%%%%%%%%%%%%%%%%%%%%%%%%%%%%%%%%%%%%%%%%%%%%%%%%%%%%%%%%%%%%%%%%%%%%%%%
			\begin{align}
			U_{1}=V^{-1}U_{1}^{\rm diag}V=\left(\begin{array}{cc}
									-1&0\\
									0&-1
									\end{array}\right),
			\end{align}
			%%%%%%%%%%%%%%%%%%%%%%%%%%%%%%%%%%%%%%%%%%%%%%%%%%%%%%%%%%%%%%%%%%%%%%%%%%
		for {Type \GkII}. Then, the condition (\ref{y1-direction-lambda-generalBC}) is trivially satisfied, while (\ref{y1-direction-rho-generalBC}) reduces to
			%%%%%%%%%%%%%%%%%%%%%%%%%%%%%%%%%%%%%%%%%%%%%%%%%%%%%%%%%%%%%%%%%%%%%%%%%%
			\begin{align}
			f_{1}(y)=f_{2}(y)=0\qquad \text{at}\ \ y_{1}=0,L_{1}.
			\end{align}
			%%%%%%%%%%%%%%%%%%%%%%%%%%%%%%%%%%%%%%%%%%%%%%%%%%%%%%%%%%%%%%%%%%%%%%%%%%
		In terms of $\Phi(y)$, the above boundary condition can be expressed as
			%%%%%%%%%%%%%%%%%%%%%%%%%%%%%%%%%%%%%%%%%%%%%%%%%%%%%%%%%%%%%%%%%%%%%%%%%%
			\begin{align}
			{\cal P}_{(-1)^{F}=+1}\Phi(y)=0\qquad \text{at}\ \ y_{1}=0,L_{1},
			\end{align}
			%%%%%%%%%%%%%%%%%%%%%%%%%%%%%%%%%%%%%%%%%%%%%%%%%%%%%%%%%%%%%%%%%%%%%%%%%%	
		for \cred{the} {Type \GkII} boundary condition.	
		
	\item[\r{3})]\ \underline{Type {\GkIII} boundary condition}\\[0.3cm]
		For {Type \GkIII}, the unitary matrix $U_{1}$ can generally be written as 
		%%%%%%%%%%%%%%%%%%%%%%%%%%%%%%%%%%%%%%%%%%%%%%%%%%%%%%%%%%%%%%%%%%%%%%%%%%
		\begin{align}
		U_{1}=V^{-1}\left(\begin{array}{cc}
								1&0\\
								0&-1
								\end{array}\right)V=V^{-1}\sigma_{3}V.\label{typeIII-BC-U1}
		\end{align}
		%%%%%%%%%%%%%%%%%%%%%%%%%%%%%%%%%%%%%%%%%%%%%%%%%%%%%%%%%%%%%%%%%%%%%%%%%%
		Since $V$ can be any element of $U(2)$, $V$ could be parameterized as
		%%%%%%%%%%%%%%%%%%%%%%%%%%%%%%%%%%%%%%%%%%%%%%%%%%%%%%%%%%%%%%%%%%%%%%%%%%
		\begin{align}
		{V=e^{ia\,{I_{2}} +ib\,\sigma_{3}}e^{i\frac{\theta_{1}}{2}(-\sin\phi_{1}\,\sigma_{1}+\cos\phi_{1}\, \sigma_{2})}.}
		\end{align}
		%%%%%%%%%%%%%%%%%%%%%%%%%%%%%%%%%%%%%%%%%%%%%%%%%%%%%%%%%%%%%%%%%%%%%%%%%%
		However, $e^{ia\,{I_{2}}+ib\,\sigma_{3}}$ trivially acts on $\sigma_{3}$ in (\ref{typeIII-BC-U1}), so that the relevant part of $V$ in the unitary transformation (\ref{typeIII-BC-U1}) will be given by
		%%%%%%%%%%%%%%%%%%%%%%%%%%%%%%%%%%%%%%%%%%%%%%%%%%%%%%%%%%%%%%%%%%%%%%%%%%
		\begin{align}
		{V=e^{i\frac{\theta_{1}}{2}(-\sin\phi_{1} \,\sigma_{1}+\cos\phi_{1} \,\sigma_{2})}.}
		\end{align}
		%%%%%%%%%%%%%%%%%%%%%%%%%%%%%%%%%%%%%%%%%%%%%%%%%%%%%%%%%%%%%%%%%%%%%%%%%%
		Then, we find \cred{that}		
		%%%%%%%%%%%%%%%%%%%%%%%%%%%%%%%%%%%%%%%%%%%%%%%%%%%%%%%%%%%%%%%%%%%%%%%%%%
		\begin{align}
		U_{1}=V^{-1}\sigma_{3}V=\vec{n}_{1}\cdot\vec{\sigma}={\left(\begin{array}{cc}
																	\cos\theta_{1}&e^{-i\phi_{1}}\sin\theta_{1}\\
																	e^{i\phi_{1}}\sin\theta_{1}&-\cos\theta_{1}
																	\end{array}\right),}\label{typeIII-BC-U1-ex}
		\end{align}
		%%%%%%%%%%%%%%%%%%%%%%%%%%%%%%%%%%%%%%%%%%%%%%%%%%%%%%%%%%%%%%%%%%%%%%%%%%
		where $\vec{\sigma}=(\sigma_{1},\sigma_{2},\sigma_{3})$ are the Pauli matrices and $\vec{n}_{1}$ is a unit vector pointing a position of a unit \cred{two-sphere} $S^2$ defined by 
		%%%%%%%%%%%%%%%%%%%%%%%%%%%%%%%%%%%%%%%%%%%%%%%%%%%%%%%%%%%%%%%%%%%%%%%%%%
		\begin{align}
		\vec{n}_{1}=(\cos\phi_{1}\sin\theta_{1},\sin\phi_{1}\sin\theta_{1},\cos\theta_{1}).
		\end{align}
		%%%%%%%%%%%%%%%%%%%%%%%%%%%%%%%%%%%%%%%%%%%%%%%%%%%%%%%%%%%%%%%%%%%%%%%%%%
		The above result shows that the parameter space of \cred{the} {Type \GkIII} boundary {condition} is given by $S^{2}=U(2)/{(U(1)\times U(1))}$. Therefore, \cred{the} {Type \GkIII} boundary condition is expected to possess rich physical implications\cred{,} because the parameter space is topologically non-trivial \cite{Cheon:2000tq}.
		
		It follows from (\ref{typeIII-BC-U1-ex}) that (\ref{y1-direction-rho-generalBC}) and (\ref{y1-direction-lambda-generalBC}) become
		%%%%%%%%%%%%%%%%%%%%%%%%%%%%%%%%%%%%%%%%%%%%%%%%%%%%%%%%%%%%%%%%%%%%%%%%%%
		\begin{align}
		&({{\rm I}_2}-\vec{n}_{1}\cdot \vec{\sigma})\left(\begin{array}{c}
												f_{1}(y)\\
												f_{2}(y)
												\end{array}\right)=0,\nonumber\\
		&({{\rm I}_2}+\vec{n}_{1}\cdot \vec{\sigma})\left(\begin{array}{c}
												-g_{2}(y)\\
												g_{1}(y)
												\end{array}\right)=0,\qquad \text{at}\ \ y_{1}=0,L_{1}. \label{typeIII-BC-rho-lambda}
		\end{align}
		%%%%%%%%%%%%%%%%%%%%%%%%%%%%%%%%%%%%%%%%%%%%%%%%%%%%%%%%%%%%%%%%%%%%%%%%%%
		It will be more convenient to express the above boundary condition in terms of the original \cred{four}-component wavefunction $\Phi(y)$. To this end, we may use the relation
		$$\left(\begin{array}{c}
		-g_{2}(y)\\
		g_{1}(y)
		\end{array}\right)=-i\sigma_{2}\left(\begin{array}{c}
		g_{1}(y)\\
		g_{2}(y)
		\end{array}\right)$$
		and combine the two conditions of (\ref{typeIII-BC-rho-lambda}) into a single one as
		%%%%%%%%%%%%%%%%%%%%%%%%%%%%%%%%%%%%%%%%%%%%%%%%%%%%%%%%%%%%%%%%%%%%%%%%%%
		\begin{align}
		{\cal P}_{\vec{n}_{1}\cdot\vec{\Sigma}_{1}=-1}\Phi(y)=0\qquad \text{at}\ \ y_{1}=0,L_{1},
		\end{align}
		%%%%%%%%%%%%%%%%%%%%%%%%%%%%%%%%%%%%%%%%%%%%%%%%%%%%%%%%%%%%%%%%%%%%%%%%%%
		where ${\cal P}_{\vec{n}_{1}\cdot\vec{\Sigma}_{1}=-1}$ is defined by
		%%%%%%%%%%%%%%%%%%%%%%%%%%%%%%%%%%%%%%%%%%%%%%%%%%%%%%%%%%%%%%%%%%%%%%%%%%
		\begin{align}
		{\cal P}_{\vec{n}_{1}\cdot\vec{\Sigma}_{1}=\pm1}&\equiv \frac{1}{2}\Bigl({\rm I}_{4}\pm\vec{n}_{1}\cdot\vec{\Sigma}_{1}\Bigr),\\
		\vec{\Sigma}_{1}&\equiv \left(\begin{array}{cc}
										\vec{\sigma}&0\\
										0&-\sigma_{2}\vec{\sigma}\sigma_{2}
										\end{array}\right).
		\end{align}
		%%%%%%%%%%%%%%%%%%%%%%%%%%%%%%%%%%%%%%%%%%%%%%%%%%%%%%%%%%%%%%%%%%%%%%%%%%	
		Since $(\vec{n}_{1}\cdot\vec{\Sigma}_{1})^{2}={\rm I}_{4}$ with $\vec{n}_{1}\cdot \vec{n}_{1}=1$, ${\cal P}_{\vec{n}_{1}\cdot\vec{\Sigma}_{1}=-1}$ can be regarded as the projection matrix on a subspace of $\vec{n}_{1}\cdot \vec{\Sigma}_{1}=-1$.
	\end{description}	
	%%%%%%%%%%%%%%%%%%%%%%%%%%%%%%%%%%%%%%%%%%%%%%%%%%%%%%%%%%%%%%%%%%%%%%%%%%	

It is interesting to note that every boundary condition of {Type} I, {\GkII}\cred{,} and {\GkIII} can be expressed by use of the projection matrices, ${\cal P}_{(-1)^{F}=-1}$, ${\cal P}_{(-1)^{F}=+1}$ and ${\cal P}_{\vec{n}\cdot\vec{\Sigma}_{1}=-1}$, respectively, and that those representations become important in the subsection \ref{sec:Verification of the sufficient condition} to verify the sufficiency of the conditions obtained above.

%%%%%%%%%%%%%%%		Subsection~4.3		%%%%%%%%%%%%%%%%%%%
\subsection{Allowed boundary conditions in the $y_{2}$-direction}
\label{sec:Allowed boundary conditions in the y_{2}-direction}

Let us next investigate the condition (\ref{y2-direction-probability-current-BC}), whose solutions will give possible boundary conditions in the $y_{2}$-direction. As before, by taking $\Phi'(y)=\Phi(y)$, (\ref{y2-direction-probability-current-BC}) is found to be written \cred{as}
	%%%%%%%%%%%%%%%%%%%%%%%%%%%%%%%%%%%%%%%%%%%%%%%%%%%%%%%%%%%%%%%%%%%%%%%%%%
	\begin{align}
	|\rho_{2}(y)+L_{0}\lambda_{2}(y)|^{2}=|\rho_{2}(y)-L_{0}\lambda_{2}(y)|^{2}\qquad \text{at}\ \ y_{2}=0,L_{2},\label{y2-ablusolute-form-rho-lambda-boundaries}
	\end{align}
	%%%%%%%%%%%%%%%%%%%%%%%%%%%%%%%%%%%%%%%%%%%%%%%%%%%%%%%%%%%%%%%%%%%%%%%%%%
where
	%%%%%%%%%%%%%%%%%%%%%%%%%%%%%%%%%%%%%%%%%%%%%%%%%%%%%%%%%%%%%%%%%%%%%%%%%%
	\begin{align}
	\rho_{2}(y)=\left(\begin{array}{c}
						f_{1}(y)\\
						f_{2}(y)
						\end{array}\right),\qquad \lambda_{2}(y)=\left(\begin{array}{c}
						g_{2}(y)\\
						g_{1}(y)
						\end{array}\right).
	\end{align}
	%%%%%%%%%%%%%%%%%%%%%%%%%%%%%%%%%%%%%%%%%%%%%%%%%%%%%%%%%%%%%%%%%%%%%%%%%%
Here, $L_{0}$ is a non-zero real constant whose value is irrelevant unless $L_{0}$ is non-vanishing. General solutions to (\ref{y2-ablusolute-form-rho-lambda-boundaries}) are given by 
	%%%%%%%%%%%%%%%%%%%%%%%%%%%%%%%%%%%%%%%%%%%%%%%%%%%%%%%%%%%%%%%%%%%%%%%%%%
	\begin{align}
	\Bigl({\rm I}_{2}-U_{2}\Bigr)\rho_{2}(y)=-L_{0}\Bigl({\rm I}_{2}+U_{2}\Bigr)\lambda_{2}(y)\qquad \text{at}\ \ y_{2}=0,L_{2},\label{y2-direction-generalBC}
	\end{align}
	%%%%%%%%%%%%%%%%%%%%%%%%%%%%%%%%%%%%%%%%%%%%%%%%%%%%%%%%%%%%%%%%%%%%%%%%%%
where $U_{2}$ is an arbitrary $2\times 2$ unitary matrix.

Requiring that the boundary conditions have to be compatible with the eigenvalues of $(-1)^{F}$, we find that (\ref{y2-direction-generalBC}) should reduce to
	%%%%%%%%%%%%%%%%%%%%%%%%%%%%%%%%%%%%%%%%%%%%%%%%%%%%%%%%%%%%%%%%%%%%%%%%%%
	\begin{align}
	&\Bigl({\rm I}_{2}-U_{2}\Bigr)\rho_{2}(y)=0,\\
	&\Bigl({\rm I}_{2}+U_{2}\Bigr)\lambda_{2}(y)=0\qquad \text{at}\ y_{2}=0,L_{2}.
	\end{align}
	%%%%%%%%%%%%%%%%%%%%%%%%%%%%%%%%%%%%%%%%%%%%%%%%%%%%%%%%%%%%%%%%%%%%%%%%%%
This implies that the eigenvalues of $U_{2}$ have to be $+1$ or $-1$. As before, we can then show that the form of $U_{2}$ is classified into three categories such as
	%%%%%%%%%%%%%%%%%%%%%%%%%%%%%%%%%%%%%%%%%%%%%%%%%%%%%%%%%%%%%%%%%%%%%%%%%%
	\begin{description}
	\item[\r{1})]\ Type I
		%%%%%%%%%%%%%%%%%%%%%%%%%%%%%%%%%%%%%%%%%%%%%%%%%%%%%%%%%%%%%%%%%%%%%%%%%%
		\begin{align}
		U_{2}={\rm I}_{2}{,}
		\end{align}
		%%%%%%%%%%%%%%%%%%%%%%%%%%%%%%%%%%%%%%%%%%%%%%%%%%%%%%%%%%%%%%%%%%%%%%%%%%
	\item[\r{2})]\ Type {\GkII}
		%%%%%%%%%%%%%%%%%%%%%%%%%%%%%%%%%%%%%%%%%%%%%%%%%%%%%%%%%%%%%%%%%%%%%%%%%%
		\begin{align}
		U_{2}=-{\rm I}_{2}{,}
		\end{align}
		%%%%%%%%%%%%%%%%%%%%%%%%%%%%%%%%%%%%%%%%%%%%%%%%%%%%%%%%%%%%%%%%%%%%%%%%%%
	\item[\r{3})]\ Type {\GkIII}
		%%%%%%%%%%%%%%%%%%%%%%%%%%%%%%%%%%%%%%%%%%%%%%%%%%%%%%%%%%%%%%%%%%%%%%%%%%
		\begin{align}
		U_{2}&=\vec{n}_{2}\cdot\vec{\sigma}=\left(\begin{array}{cc}
													\cos\theta_{2}&e^{-i\phi_{2}}\sin\theta_{2}\\
													e^{i\phi_{2}}\sin\theta_{2}&-\cos\theta_{2}
													\end{array}\right),\nonumber\\
		\vec{n}_{2}&=(\cos\phi_{2}\sin\theta_{2},\sin\phi_{2}\sin\theta_{2},\cos\theta_{2}).													
		\end{align}
		%%%%%%%%%%%%%%%%%%%%%%%%%%%%%%%%%%%%%%%%%%%%%%%%%%%%%%%%%%%%%%%%%%%%%%%%%%
	\end{description}
	%%%%%%%%%%%%%%%%%%%%%%%%%%%%%%%%%%%%%%%%%%%%%%%%%%%%%%%%%%%%%%%%%%%%%%%%%%
It follows that allowed boundary conditions are given by
	%%%%%%%%%%%%%%%%%%%%%%%%%%%%%%%%%%%%%%%%%%%%%%%%%%%%%%%%%%%%%%%%%%%%%%%%%%
	\begin{description}
	\item[\r{1})]\ Type I boundary condition
		%%%%%%%%%%%%%%%%%%%%%%%%%%%%%%%%%%%%%%%%%%%%%%%%%%%%%%%%%%%%%%%%%%%%%%%%%%
		\begin{align}
		{\cal P}_{(-1)^{F}=-1}\Phi(y)=0\qquad \text{at}\ \ y_{2}=0,L_{2}.
		\end{align}
		%%%%%%%%%%%%%%%%%%%%%%%%%%%%%%%%%%%%%%%%%%%%%%%%%%%%%%%%%%%%%%%%%%%%%%%%%%
	\item[\r{2})]\ Type {\GkII} boundary condition
		%%%%%%%%%%%%%%%%%%%%%%%%%%%%%%%%%%%%%%%%%%%%%%%%%%%%%%%%%%%%%%%%%%%%%%%%%%
		\begin{align}
		{\cal P}_{(-1)^{F}=+1}\Phi(y)=0\qquad \text{at}\ \ y_{2}=0,L_{2}.
		\end{align}
		%%%%%%%%%%%%%%%%%%%%%%%%%%%%%%%%%%%%%%%%%%%%%%%%%%%%%%%%%%%%%%%%%%%%%%%%%%
	\item[\r{3})]\ Type {\GkIII} boundary condition
		%%%%%%%%%%%%%%%%%%%%%%%%%%%%%%%%%%%%%%%%%%%%%%%%%%%%%%%%%%%%%%%%%%%%%%%%%%
		\begin{align}
		{\cal P}_{\vec{n}_{2}\cdot\vec{\Sigma}_{2}=-1}\Phi(y)=0\qquad \text{at}\ \ y_{2}=0,L_{2},
		\end{align}
		%%%%%%%%%%%%%%%%%%%%%%%%%%%%%%%%%%%%%%%%%%%%%%%%%%%%%%%%%%%%%%%%%%%%%%%%%%
\hspace{-1em}where ${\cal P}_{\vec{n}_{2}\cdot\vec{\Sigma}_{2}=-1}$ is a projection matrix defined by
		%%%%%%%%%%%%%%%%%%%%%%%%%%%%%%%%%%%%%%%%%%%%%%%%%%%%%%%%%%%%%%%%%%%%%%%%%%
		\begin{align}
		&{\cal P}_{\vec{n}_{2}\cdot\vec{\Sigma}_{2}=\pm1}\equiv \frac{1}{2}({\rm I}_{4}\pm\vec{n}_{2}\cdot\vec{\Sigma}_{2}),\\
		&\vec{\Sigma}_{2}\equiv \left(\begin{array}{cc}
										\vec{\sigma}&0\\
										0&-\sigma_{1}\vec{\sigma}\sigma_{1}
										\end{array}\right).
		\end{align}
		%%%%%%%%%%%%%%%%%%%%%%%%%%%%%%%%%%%%%%%%%%%%%%%%%%%%%%%%%%%%%%%%%%%%%%%%%%	
	\end{description}
	%%%%%%%%%%%%%%%%%%%%%%%%%%%%%%%%%%%%%%%%%%%%%%%%%%%%%%%%%%%%%%%%%%%%%%%%%%

%%%%%%%%%%%%%%%		Subsection~4.4		%%%%%%%%%%%%%%%%%%%
\subsection{Verification of the sufficient condition}
\label{sec:Verification of the sufficient condition}

We have succeeded in classifying the allowed boundary conditions into three categories that satisfy (\ref{y1-direction-probability-current-BC}) or (\ref{y2-direction-probability-current-BC}) with the restriction of $\Phi'(y) =\Phi(y)$. In the following, we show that the boundary conditions derived in the subsection \ref{sec:Allowed boundary conditions in the y_{1}-direction} and \ref{sec:Allowed boundary conditions in the y_{2}-direction} \cred{in fact} satisfy (\ref{y1-direction-probability-current-BC}) and (\ref{y2-direction-probability-current-BC}) even for independent $\Phi(y)$ and $\Phi'(y)$. For our purpose, it will be convenient to rewrite (\ref{y1-direction-probability-current-BC}) and (\ref{y2-direction-probability-current-BC}) into the form
	%%%%%%%%%%%%%%%%%%%%%%%%%%%%%%%%%%%%%%%%%%%%%%%%%%%%%%%%%%%%%%%%%%%%%%%%%%
	\begin{align}
	\Bigl(\Phi'(y)\Bigr)^{\dagger}\widetilde{\Gamma}_{1}\Phi(y)=0\qquad \text{at}\ \ y_{1}=0,L_{1},\label{Phi-y1-direction-BC}\\
	\Bigl(\Phi'(y)\Bigr)^{\dagger}\widetilde{\Gamma}_{2}\Phi(y)=0\qquad \text{at}\ \ y_{2}=0,L_{2},\label{Phi-y2-direction-BC}
	\end{align}
	%%%%%%%%%%%%%%%%%%%%%%%%%%%%%%%%%%%%%%%%%%%%%%%%%%%%%%%%%%%%%%%%%%%%%%%%%%
where
	%%%%%%%%%%%%%%%%%%%%%%%%%%%%%%%%%%%%%%%%%%%%%%%%%%%%%%%%%%%%%%%%%%%%%%%%%%
	\begin{align}
	\widetilde{\Gamma}_{1}\equiv \left(\begin{array}{cc}
										0&{-\sigma_{2}}\\
										{-\sigma_{2}}&0
										\end{array}\right),\qquad \widetilde{\Gamma}_{2}\equiv \left(\begin{array}{cc}
										0&\sigma_{1}\\
										\sigma_{1}&0
										\end{array}\right).
	\end{align}
	%%%%%%%%%%%%%%%%%%%%%%%%%%%%%%%%%%%%%%%%%%%%%%%%%%%%%%%%%%%%%%%%%%%%%%%%%%
	%%%%%%%%%%%%%%%%%%%%%%%%%%%%%%%%%%%%%%%%%%%%%%%%%%%%%%%%%%%%%%%%%%%%%%%%%%
	\begin{description}
	\item[\r{1})] \ \underline{Type I {boundary condition} in the $y_{1}$-direction}\\[0.3cm]
		We first investigate \cred{the} {Type} I boundary condition in the $y_{1}$-direction, i.e.
			%%%%%%%%%%%%%%%%%%%%%%%%%%%%%%%%%%%%%%%%%%%%%%%%%%%%%%%%%%%%%%%%%%%%%%%%%%
			\begin{align}
			{\cal P}_{(-1)^{F}=-1}\Phi(y)={\cal P}_{(-1)^{F}=-1}\Phi'(y)=0\qquad \text{at}\ \ y_{1}=0,L_{1}. \label{sufficient-y1-Phi-Phidach}
			\end{align}
			%%%%%%%%%%%%%%%%%%%%%%%%%%%%%%%%%%%%%%%%%%%%%%%%%%%%%%%%%%%%%%%%%%%%%%%%%%
		Important properties \cred{for proving} the condition (\ref{Phi-y1-direction-BC}) are
		%%%%%%%%%%%%%%%%%%%%%%%%%%%%%%%%%%%%%%%%%%%%%%%%%%%%%%%%%%%%%%%%%%%%%%%%%%
		\begin{align}
		&{\cal P}_{(-1)^{F}=+1}+{\cal P}_{(-1)^{F}=-1}={\rm I}_{4},\nonumber\\
		&\Bigl({\cal P}_{(-1)^{F}=\pm 1}\Bigr)^{2}={\cal P}_{(-1)^{F}=\pm 1},\qquad {\cal P}_{(-1)^{F}=\pm 1}{\cal P}_{(-1)^{F}=\mp 1}=0,\nonumber\\
		&\Bigl({\cal P}_{(-1)^{F}=\pm 1}\Bigr)^{\dagger}={\cal P}_{(-1)^{F}=\pm 1},\nonumber\\
		&{\cal P}_{(-1)^{F}=\pm 1}\widetilde{\Gamma}_{1}=\widetilde{\Gamma}_{1}{\cal P}_{(-1)^{F}=\mp 1},\label{y1-projection-operator}
		\end{align}
		%%%%%%%%%%%%%%%%%%%%%%%%%%%%%%%%%%%%%%%%%%%%%%%%%%%%%%%%%%%%%%%%%%%%%%%%%%
		where the last relation follows from $(-1)^{F}\,\widetilde{\Gamma}_{1}=-\widetilde{\Gamma}_{1}(-1)^{F}$. With a shorthand notation of $\Phi_{\pm}(y)\equiv {\cal P}_{(-1)^{F}=\pm 1}\Phi(y)$, the condition (\ref{Phi-y1-direction-BC}) can be verified as follows:
		%%%%%%%%%%%%%%%%%%%%%%%%%%%%%%%%%%%%%%%%%%%%%%%%%%%%%%%%%%%%%%%%%%%%%%%%%%
		\begin{align}
		\Bigl(\Phi'(y)\Bigr)^{\dagger}\widetilde{\Gamma}_{1}\Phi(y)&=\Bigl(\Phi'_{+}(y)+\Phi'_{-}(y)\Bigr)^{\dagger}\widetilde{\Gamma}_{1}\Bigl(\Phi_{+}(y)+\Phi_{-}(y)\Bigr)\nonumber\\
		&=\Bigl(\Phi'_{+}(y)\Bigr)^{\dagger}\widetilde{\Gamma}_{1}\Phi_{-}(y)+\Bigl(\Phi'_{-}(y)\Bigr)^{\dagger}\widetilde{\Gamma}_{1}\Phi_{+}(y)\nonumber\\
		&=0\qquad \text{at}\ \ y_{1}=0,L_{1},
		\end{align}
		%%%%%%%%%%%%%%%%%%%%%%%%%%%%%%%%%%%%%%%%%%%%%%%%%%%%%%%%%%%%%%%%%%%%%%%%%%
		where we have used the relations (\ref{sufficient-y1-Phi-Phidach}) and (\ref{y1-projection-operator}).
		
		\item[\r{2})] \ \underline{Type {\GkII} {boundary condition} in the $y_{1}$-direction}\\[0.3cm]
		The above analysis for \cred{the} {Type} I boundary condition clearly shows that if $\Phi'(y)$ and $\Phi(y)$ satisfy the {Type \GkII} boundary condition in the $y_{1}$-direction, i.e.
		%%%%%%%%%%%%%%%%%%%%%%%%%%%%%%%%%%%%%%%%%%%%%%%%%%%%%%%%%%%%%%%%%%%%%%%%%%
		\begin{align}
		{\cal P}_{(-1)^{F}=+1}\Phi'(y)={\cal P}_{(-1)^{F}=+1}\Phi(y)=0\qquad \text{at}\ \ y_{1}=0,L_{1},\label{typeII-BC-projection-form}
		\end{align}
		%%%%%%%%%%%%%%%%%%%%%%%%%%%%%%%%%%%%%%%%%%%%%%%%%%%%%%%%%%%%%%%%%%%%%%%%%%
		then the condition (\ref{Phi-y1-direction-BC}) is satisfied for arbitrary wavefunctions $\Phi(y)$ and $\Phi'(y)$ with (\ref{typeII-BC-projection-form}).
		
		\item[\r{3})] \ \underline{Type {\GkIII} {boundary condition} in the $y_{1}$-direction}\\[0.3cm]
		In order to prove that \cred{the} {Type \GkIII} boundary condition in the $y_{1}$-direction satisfies the condition (\ref{Phi-y1-direction-BC}), we need the following properties of ${\cal P}_{\vec{n}_{1}\cdot\vec{\Sigma}_{1}=\pm 1}$:
		%%%%%%%%%%%%%%%%%%%%%%%%%%%%%%%%%%%%%%%%%%%%%%%%%%%%%%%%%%%%%%%%%%%%%%%%%%
		\begin{align}
		&{\cal P}_{\vec{n}_{1}\cdot\vec{\Sigma}_{1}=+1}+{\cal P}_{\vec{n}_{1}\cdot\vec{\Sigma}_{1}=-1} ={\rm I}_{4},\nonumber\\
		&\Bigl({\cal P}_{\vec{n}\cdot \vec{\Sigma}_{1}=\pm 1}\Bigr)^{2}={\cal P}_{\vec{n}\cdot \vec{\Sigma}_{1}=\pm 1},\qquad {\cal P}_{\vec{n}\cdot \vec{\Sigma}_{1}=\pm 1}{\cal P}_{\vec{n}\cdot \vec{\Sigma}_{1}=\mp 1}=0,\nonumber\\
		&\Bigl({\cal P}_{\vec{n}\cdot \vec{\Sigma}_{1}=\pm 1}\Bigr)^{\dagger}={\cal P}_{\vec{n}\cdot \vec{\Sigma}_{1}=\pm 1},\nonumber\\
		&{\cal P}_{\vec{n}\cdot \vec{\Sigma}_{1}=\pm 1}\widetilde{\Gamma}_{1}=\widetilde{\Gamma}_{1}{\cal P}_{\vec{n}\cdot \vec{\Sigma}_{1}=\mp 1},
		\end{align}
		%%%%%%%%%%%%%%%%%%%%%%%%%%%%%%%%%%%%%%%%%%%%%%%%%%%%%%%%%%%%%%%%%%%%%%%%%%
		where the last relation follows from the property \cred{$\vec{\Sigma}_{1}\widetilde{\Gamma}_{1}=-\widetilde{\Gamma}_{1}\vec{\Sigma}_{1}$}. The above relations are enough to show that if $\Phi(y)$ and $\Phi'(y)$ obey \cred{the}
		{Type \GkIII} boundary condition in the $y_{1}$-direction, they satisfy the condition (\ref{Phi-y1-direction-BC}).
	\end{description}		
	%%%%%%%%%%%%%%%%%%%%%%%%%%%%%%%%%%%%%%%%%%%%%%%%%%%%%%%%%%%%%%%%%%%%%%%%%%

The above analysis can also apply {to} {Type} I, {\GkII}\cred{,} and {\GkIII} boundary conditions in the $y_{2}$-direction. In order to verify the condition (\ref{Phi-y2-direction-BC}) for {Type} I, {\GkII}\cred{,} and {\GkIII} in the $y_{2}$-direction, we only need the properties that ${\cal P}_{(-1)^{F}=\pm 1}$ and ${\cal P}_{\vec{n}_{2}\cdot\vec{\Sigma}_{2}=\pm 1}$ can be regarded as projection matrices and that $\widetilde{\Gamma}_{2}$ changes the sign of the eigenvalues of $(-1)^{F}$ and $\vec{n}_{2}\cdot\vec{\Sigma}_{2}$. The proof can be done in a similar way as the case of $y_{1}$-direction.

%%%%%%%%%%%%%%%%%%%%%%%%%%%%%%%%%%%%%
%%%%%%%% Section 5 %%%%%%%%%%%%%%%%%%
%%%%%%%%%%%%%%%%%%%%%%%%%%%%%%%%%%%%%
\section{Energy spectrum for {Type \GkII} \cred{boundary conditions}
\label{sec:Energy spectrum for type II BC}}

In this section, we investigate the energy spectrum of the theory for {Type \GkII} boundary condition with the help of \cred{supersymmetry}.\footnote{The analysis for \cred{the} {Type} I boundary condition is almost the same as that for \cred{{Type \GkII}}.}
We will show that {Type \GkII} boundary condition is enough to determine the \cred{positive-energy spectrum completely,} but not to determine \cred{zero-energy} solutions.

%%%%%%%%%%%%%%%		Subsection~5.1		%%%%%%%%%%%%%%%%%%%
\subsection{\cred{Supersymmetry} relations and boundary conditions}
\label{sec:SUSY relations and boundary conditions}

In this subsection, we summarize \cred{the} general properties of $N=2$ supersymmetric quantum mechanics to determine the energy spectrum.
	
Let $\Phi_{E\pm}(y)$ be simultaneous eigenstates of $H$ and $(-1)^{F}$, i.e.
	%%%%%%%%%%%%%%%%%%%%%%%%%%%%%%%%%%%%%%%%%%%%%%%%%%%%%%%%%%%%%%%%%%%%%%%%%%
	\begin{align}
	H\Phi_{E\pm}(y)&=E\Phi_{E\pm}(y),\\
	(-1)^{F}\Phi_{E\pm}(y)&=\pm\Phi_{E\pm}(y).\label{-1F-eigenstates}
	\end{align}
	%%%%%%%%%%%%%%%%%%%%%%%%%%%%%%%%%%%%%%%%%%%%%%%%%%%%%%%%%%%%%%%%%%%%%%%%%%
Since the supercharge $Q$ commutes with $H$ and anticommutes with $(-1)^{F}$, $Q\Phi_{E\pm}$ turns out to have the same energy $E$ but opposite {eigenvalues} of $(-1)^{F}$ if $Q\Phi_{E\pm}$ are non-vanishing. This implies that $Q\Phi_{E\pm}$ should be proportional to $\Phi_{E\mp}$,\footnote{If the energy spectrum {has another kind of degeneracy}, we may replace $\Phi_{E\pm}$ by $\Phi^{(i)}_{E\pm}$ with the index $i$ to distinguish degenerate states.} i.e.
	%%%%%%%%%%%%%%%%%%%%%%%%%%%%%%%%%%%%%%%%%%%%%%%%%%%%%%%%%%%%%%%%%%%%%%%%%%
	\begin{align}
	Q\Phi_{E+}(y)=\sqrt{E}\Phi_{E-}(y),\label{SUSYrelation+-}\\
	Q\Phi_{E-}(y)=\sqrt{E}\Phi_{E+}(y).\label{SUSYrelation-+}
	\end{align}
	%%%%%%%%%%%%%%%%%%%%%%%%%%%%%%%%%%%%%%%%%%%%%%%%%%%%%%%%%%%%%%%%%%%%%%%%%%
Then, $\{\Phi_{E+},\Phi_{E-}\}$ turns out to form a supermultiplet (for $E>0$), and (\ref{SUSYrelation+-}), (\ref{SUSYrelation-+}) are called the supersymmetry relations or simply SUSY relations. The factor $\sqrt{E}$ on the right-hand-sides \cred{ensures} that $\langle \Phi_{E+},\Phi_{E+}\rangle =\langle \Phi_{E-},\Phi_{E-}\rangle$.

We should emphasize that \cred{zero-energy} solutions with $E=0$ do not form supermultiplets, as suggested by the SUSY relations \cred{because zero-energy} solutions have to satisfy the \cred{zero-energy} equation\,\footnote{Since the Hamiltonian takes the form $H=Q^{2}$, the equation $H\Phi_{E=0}=0$ becomes identical to $Q\Phi_{E=0}=0$.}
	%%%%%%%%%%%%%%%%%%%%%%%%%%%%%%%%%%%%%%%%%
	%%%%%%%%%%%%%%%%%%%%%%%%%%%%%%%%%%%%%%%%%%%%%%%%%%%%%%%%%%%%%%%%%%%%%%%%%%
	\begin{align}
	Q\Phi_{E=0}(y)=0.
	\end{align}
	%%%%%%%%%%%%%%%%%%%%%%%%%%%%%%%%%%%%%%%%%%%%%%%%%%%%%%%%%%%%%%%%%%%%%%%%%%
In this section, we impose \cred{the} {Type \GkII} boundary condition on the wavefunction $\Phi(y)$ in \cred{both} the $y_{1}$- and {$y_{2}$-directions}, i.e.
	%%%%%%%%%%%%%%%%%%%%%%%%%%%%%%%%%%%%%%%%%%%%%%%%%%%%%%%%%%%%%%%%%%%%%%%%%%
	\begin{align}
	\Phi_{+}(y)=0 \qquad \text{at}\ \ y_{1}=0,L_{1}\ \ \text{and}\ \ y_{2}=0,L_{2}.\label{typeII-BC-y1-y2}
	\end{align}
	%%%%%%%%%%%%%%%%%%%%%%%%%%%%%%%%%%%%%%%%%%%%%%%%%%%%%%%%%%%%%%%%%%%%%%%%%%
One might think that (\ref{typeII-BC-y1-y2}) is not enough to specify the boundary condition for all the components of $\Phi (y)$ because (\ref{typeII-BC-y1-y2}) seems to give no constraint on $\Phi_{-}(y)$ at the boundaries. This is{, however, not} the case. The boundary condition for $\Phi_{-}(y)$ can be obtained through the SUSY relation (\ref{SUSYrelation-+}). In order for the boundary condition (\ref{typeII-BC-y1-y2}) to be consistent with the SUSY relation (\ref{SUSYrelation-+}), the wavefunction $\Phi_{-}(y)$ with $(-1)^{F}=-1$ has to obey the following boundary condition\,\footnote{The same situation has been observed in the 5d fermion system on an interval \cite{Fujimoto:2011kf, Fujimoto:2012wv, Fujimoto:2014fka} and also in supersymmetric quantum mechanics with boundaries \cite{Cheon:2000tq, Nagasawa:2008an}.}
	%%%%%%%%%%%%%%%%%%%%%%%%%%%%%%%%%%%%%%%%%
	%%%%%%%%%%%%%%%%%%%%%%%%%%%%%%%%%%%%%%%%%%%%%%%%%%%%%%%%%%%%%%%%%%%%%%%%%%
	\begin{align}
	Q\Phi_{-}(y)=0\qquad \text{at}\ \ y_{1}=0,L_{1}\ \ \text{and}\ \ y_{2}=0,L_{2}, \label{SUSY-boundaries--}
	\end{align}
	%%%%%%%%%%%%%%%%%%%%%%%%%%%%%%%%%%%%%%%%%%%%%%%%%%%%%%%%%%%%%%%%%%%%%%%%%%
otherwise the supersymmetry would be lost due to the \cred{breakdown} of the SUSY relation (\ref{SUSYrelation-+}). As we will see in the next subsection, the boundary conditions (\ref{typeII-BC-y1-y2}) and (\ref{SUSY-boundaries--}) work well to determine the \cred{positive-energy} spectrum.

%%%%%%%%%%%%%%%		Subsection~5.2	%%%%%%%%%%%%%%%%%%%
\subsection{\cred{Positive-energy} spectrum}
\label{sec:Positive energy spectrum}

In the following, we clarify the \cred{positive-energy} spectrum for \cred{the} {Type \GkII} boundary condition with the help of the supersymmetry.

In terms of the component fields $\Phi(y)=(f_{1}(y),f_{2}(y),g_{1}(y),g_{2}(y))^{\rm T}$, the \cred{Type} {\GkII} boundary condition (\ref{typeII-BC-y1-y2}) for $f_{1}(y)$ and $f_{2}(y)$ is given by
	%%%%%%%%%%%%%%%%%%%%%%%%%%%%%%%%%%%%%%%%%%%%%%%%%%%%%%%%%%%%%%%%%%%%%%%%%%
	\begin{align}
	f_{1}(y)=f_{2}(y)=0\qquad \text{at}\ \ y_{1}=0,L_{1}\ \ \text{and}\ \ y_{2}=0,L_{2}\label{typeII-BC-f1f2}
	\end{align}
	%%%%%%%%%%%%%%%%%%%%%%%%%%%%%%%%%%%%%%%%%%%%%%%%%%%%%%%%%%%%%%%%%%%%%%%%%%
and the boundary condition (\ref{SUSY-boundaries--}) for $g_{1}(y)$ and $g_{2}(y)$ is given by
	%%%%%%%%%%%%%%%%%%%%%%%%%%%%%%%%%%%%%%%%%%%%%%%%%%%%%%%%%%%%%%%%%%%%%%%%%%
	\begin{align}
	\begin{array}{l}
	Mg_{1}(y)-(\partial_{y_{1}}-i\partial_{y_{2}})g_{2}(y)=0,\\[0.2cm]
	(\partial_{y_{1}}+i\partial_{y_{2}})g_{1}(y)-Mg_{2}(y)=0,
	\end{array}\qquad \text{at}\ \ y_{1}=0,L_{1}\ \ \text{and}\ \ y_{2}=0,L_{2}. \label{typeII-BC-g1g2}
	\end{align}
	%%%%%%%%%%%%%%%%%%%%%%%%%%%%%%%%%%%%%%%%%%%%%%%%%%%%%%%%%%%%%%%%%%%%%%%%%%
	
Let $\Phi_{E+}(y)$ be an energy eigenstate with $(-1)^{F}=+1$. In components, the relation $H\Phi_{E+}(y)=E\Phi_{E+}(y)$ is rewritten as
	%%%%%%%%%%%%%%%%%%%%%%%%%%%%%%%%%%%%%%%%%%%%%%%%%%%%%%%%%%%%%%%%%%%%%%%%%%
	\begin{align}
	\Bigl[-(\partial_{y_{1}})^{2}-(\partial_{y_{2}})^{2}+M^{2}\Bigr]\left(\begin{array}{c}
												f_{1E}(y)\\
												f_{2E}(y)
												\end{array}\right)=E\left(\begin{array}{c}
												f_{1E}(y)\\
												f_{2E}(y)
												\end{array}\right).
	\end{align}
	%%%%%%%%%%%%%%%%%%%%%%%%%%%%%%%%%%%%%%%%%%%%%%%%%%%%%%%%%%%%%%%%%%%%%%%%%%
Then, the energy eigenfunctions satisfying \cred{the} {Type \GkII}  boundary condition (\ref{typeII-BC-f1f2}) are easily found to be of the form
	%%%%%%%%%%%%%%%%%%%%%%%%%%%%%%%%%%%%%%%%%%%%%%%%%%%%%%%%%%%%%%%%%%%%%%%%%%
	\begin{align}
	\Phi^{(1)}_{E_{n_{1}n_{2}+}}(y)=\left(\begin{array}{c}
												f_{n_{1}n_{2}}(y)\\
												0\\
												0\\
												0
												\end{array}\right),\qquad \Phi^{(2)}_{E_{n_{1}n_{2}+}}(y)=\left(\begin{array}{c}
												0\\
												f_{n_{1}n_{2}}(y)\\
												0\\
												0
												\end{array}\right),\label{Phi+energy-eigenfunctions}
	\end{align}
	%%%%%%%%%%%%%%%%%%%%%%%%%%%%%%%%%%%%%%%%%%%%%%%%%%%%%%%%%%%%%%%%%%%%%%%%%%
where
	%%%%%%%%%%%%%%%%%%%%%%%%%%%%%%%%%%%%%%%%%%%%%%%%%%%%%%%%%%%%%%%%%%%%%%%%%%
	\begin{align}
	&f_{n_{1}n_{2}}(y)=\frac{2}{\sqrt{L_{1}L_{2}}}\sin\Bigl(\frac{n_{1}\pi}{L_{1}}y_{1}\Bigr)\sin\Bigl(\frac{n_{2}\pi}{L_{2}}y_{2}\Bigr),\label{fn1n2}\\
	&E_{n_{1}n_{2}}=\Bigl(\frac{n_{1}\pi}{L_{1}}\Bigr)^{2}+\Bigl(\frac{n_{2}\pi}{L_{2}}\Bigr)^{2}+M^{2}
	\end{align}
	%%%%%%%%%%%%%%%%%%%%%%%%%%%%%%%%%%%%%%%%%%%%%%%%%%%%%%%%%%%%%%%%%%%%%%%%%%
for $n_{1}, n_{2} =1,2,3,\cred{\ldots}$
The eigenfunctions $f_{n_{1}n_{2}}(y)$ satisfy
	%%%%%%%%%%%%%%%%%%%%%%%%%%%%%%%%%%%%%%%%%%%%%%%%%%%%%%%%%%%%%%%%%%%%%%%%%%
	\begin{align}
	&\langle f_{m_{1}m_{2}},f_{n_{1}n_{2}}\rangle =\delta_{m_{1},n_{1}}\delta_{m_{2},n_{2}},\\
	&f_{n_{1}n_{2}}(y)=0\qquad \text{at}\ y_{1}=0, L_{1}\ \text{and}\ y_{2}=0,L_{2}
	\end{align}
	%%%%%%%%%%%%%%%%%%%%%%%%%%%%%%%%%%%%%%%%%%%%%%%%%%%%%%%%%%%%%%%%%%%%%%%%%%
for $m_{1}, m_{2}, n_{1}, n_{2}=1,2,3,\cred{\ldots}$
It should be noticed that the energy eigenfunctions (\ref{Phi+energy-eigenfunctions}) give a complete set of the function $\Phi_{+}(y)$, since the set of $\{f_{n_{1}n_{2}}(y);\, n_{1},n_{2}=1,2,3,\cred{\ldots}\}$ forms a complete set of the function satisfying the boundary condition $f(y)=0$ at $y_{1}=0,L_{1}$ and $y_{2}=0,L_{2}$.

As was explained in the subsection \ref{sec:SUSY relations and boundary conditions}, the energy eigenfunctions for $\Phi_{E-}$ can be obtained through the SUSY relation (\ref{SUSYrelation+-}), i.e.
	%%%%%%%%%%%%%%%%%%%%%%%%%%%%%%%%%%%%%%%%%%%%%%%%%%%%%%%%%%%%%%%%%%%%%%%%%%
	\begin{align}
	&\Phi^{(1)}_{E_{n_{1}n_{2}-}}(y)=\frac{1}{\sqrt{E_{n_{1}n_{2}}}}Q\Phi^{(1)}_{E_{n_{1}n_{2}+}}(y)=\frac{1}{\sqrt{E_{n_{1}n_{2}}}}\left(\begin{array}{c}
												0\\
												0\\
												Mf_{n_{1}n_{2}}(y)\\
												(\partial_{y_{1}}+i\partial_{y_{2}})f_{n_{1}n_{2}}(y)
												\end{array}\right),\nonumber\\
	&\Phi^{(2)}_{E_{n_{1}n_{2}-}}(y)=\frac{1}{\sqrt{E_{n_{1}n_{2}}}}Q\Phi^{(2)}_{E_{n_{1}n_{2}+}}(y)=\frac{1}{\sqrt{E_{n_{1}n_{2}}}}\left(\begin{array}{c}
												0\\
												0\\
												-(\partial_{y_{1}}-i\partial_{y_{2}})f_{n_{1}n_{2}}(y)\\
												-M f_{n_{1}n_{2}}(y)
												\end{array}\right),										
	\end{align}
	%%%%%%%%%%%%%%%%%%%%%%%%%%%%%%%%%%%%%%%%%%%%%%%%%%%%%%%%%%%%%%%%%%%%%%%%%%
except for \cred{zero-energy} solutions. We note that the above eigenfunctions satisfy the boundary conditions (\ref{SUSY-boundaries--}) or (\ref{typeII-BC-g1g2}), as they \cred{should}.

%%%%%%%%%%%%%%%		Subsection~5.3		%%%%%%%%%%%%%%%%%%%
\subsection{\cred{Zero-energy} solutions}
\label{sec:Zero energy solutions}

In the previous analysis, we have succeeded in constructing \cred{positive-energy} solutions, completely. The analysis is, however, insufficient to obtain the whole set of \cred{energy eigenfunctions}.
This is because \cred{zero-energy} solutions do not form supermultiplets and hence we have to investigate them \cred{separately}.

As was explained in the subsection \ref{sec:SUSY relations and boundary conditions}, any \cred{zero-energy} solution should satisfy the \cred{zero-energy} equation $Q\Phi_{E=0}(y)=0$. Since $\Phi_{+}(y)$ has no \cred{zero-energy} solution due to the Dirichlet boundary condition (i.e. {Type \GkII} boundary condition), \cred{zero-energy} eigenfunctions will come only from $\Phi_{-}(y)$ (or $g_{1}(y)$ and $g_{2}(y)$) satisfying $Q\Phi_{E=0-}(y)=0$, or in components
	%%%%%%%%%%%%%%%%%%%%%%%%%%%%%%%%%%%%%%%%%%%%%%%%%%%%%%%%%%%%%%%%%%%%%%%%%%
	\begin{align}
	&Mg_{1E=0}(y)-(\partial_{y_{1}}-i\partial_{y_{2}})g_{2E=0}(y)=0,\nonumber\\
	&(\partial_{y_{1}}+i\partial_{y_{2}})g_{1E=0}(y)-Mg_{2E=0}(y)=0.\label{zero-energy-equations-g1g2}
	\end{align}
	%%%%%%%%%%%%%%%%%%%%%%%%%%%%%%%%%%%%%%%%%%%%%%%%%%%%%%%%%%%%%%%%%%%%%%%%%%
It is worth while pointing out that a strange situation happens here. We have already found that the boundary condition (\ref{typeII-BC-g1g2}) for $g_{1}(y)$ and $g_{2}(y)$ works properly for \cred{positive-energy} eigenstates. The boundary condition (\ref{typeII-BC-g1g2}), however, gives {\it no} restriction on \cred{zero-energy} solutions because any \cred{zero-energy} solutions {\it trivially} satisfy the ``boundary condition'' (\ref{typeII-BC-g1g2}) not only at the boundaries but also on the whole space of the rectangle. {In fact,} the condition (\ref{typeII-BC-g1g2}) can be regarded as \cred{part of the zero-energy} equation (\ref{zero-energy-equations-g1g2}).\footnote{A similar situation has been observed in the 5d fermion system on an interval \cite{Fujimoto:2011kf, Fujimoto:2012wv, Fujimoto:2014fka}.}
This implies that the determination of \cred{zero-energy} solutions might be ambiguous, as we will see below.
	
A \cred{zero-energy} solution to (\ref{zero-energy-equations-g1g2}) is found to be of the form
	%%%%%%%%%%%%%%%%%%%%%%%%%%%%%%%%%%%%%%%%%%%%%%%%%%%%%%%%%%%%%%%%%%%%%%%%%%
	\begin{align}
	\Phi^{(1)}_{E=0-}(y)=\left(\begin{array}{c}
						0\\
						0\\
						Ne^{-i\frac{\theta}{2}}e^{M(\cos\theta\,y_{1}+\sin\theta\,y_{2})}\\
						Ne^{i\frac{\theta}{2}}e^{M(\cos\theta\,y_{1}+\sin\theta\,y_{2})}
						\end{array}\right), \label{zero-mode-solution-1-}
	\end{align}
	%%%%%%%%%%%%%%%%%%%%%%%%%%%%%%%%%%%%%%%%%%%%%%%%%%%%%%%%%%%%%%%%%%%%%%%%%%
where $\theta$ is an arbitrary real constant\,\footnote{It has been shown in Ref.~{\cite{Fujimoto:2016llj}} that the origin of the parameter $\theta$ in (\ref{zero-mode-solution-1-}) comes from the rotational invariance of the extra dimensions.}
	%%%%%%%%%%%%%%%%%%%%%%%%%%%%%%%%%%%%%%%%%
and $N$ stands for a normalization constant. We will \cred{comment} {on general \cred{zero-energy} solutions} later.

We would like to know how many independent \cred{zero-energy} solutions exist in the model. To this end, we may assume \cred{a} second \cred{zero-energy} solution to be of the form
	%%%%%%%%%%%%%%%%%%%%%%%%%%%%%%%%%%%%%%%%%%%%%%%%%%%%%%%%%%%%%%%%%%%%%%%%%%
	\begin{align}
		\Phi^{(2)}_{E=0-}(y)=\left(\begin{array}{c}
						0\\
						0\\
						N'e^{-i\frac{\theta'}{2}}e^{M(\cos\theta'\,y_{1}+\sin\theta'\,y_{2})}\\
						N'e^{i\frac{\theta'}{2}}e^{M(\cos\theta'\,y_{1}+\sin\theta'\,y_{2})}
						\end{array}\right).\label{another-type-of-zero-energy-solution}
	\end{align}
	%%%%%%%%%%%%%%%%%%%%%%%%%%%%%%%%%%%%%%%%%%%%%%%%%%%%%%%%%%%%%%%%%%%%%%%%%%
In order for $\Phi^{(1)}_{E=0-}$ and $\Phi^{(2)}_{E=0-}$ to be independent, we require that they are orthogonal, i.e.
	%%%%%%%%%%%%%%%%%%%%%%%%%%%%%%%%%%%%%%%%%%%%%%%%%%%%%%%%%%%%%%%%%%%%%%%%%%
	\begin{align}
	\langle \Phi^{(1)}_{E=0-},\Phi^{(2)}_{E=0-}\rangle=0.
	\end{align}
	%%%%%%%%%%%%%%%%%%%%%%%%%%%%%%%%%%%%%%%%%%%%%%%%%%%%%%%%%%%%%%%%%%%%%%%%%%
It follows that the above orthogonality relation is satisfied only if
	%%%%%%%%%%%%%%%%%%%%%%%%%%%%%%%%%%%%%%%%%%%%%%%%%%%%%%%%%%%%%%%%%%%%%%%%%%
	\begin{align}
	\theta' =\theta +\pi\qquad (\text{mod}\ 2\pi).
	\end{align}
	%%%%%%%%%%%%%%%%%%%%%%%%%%%%%%%%%%%%%%%%%%%%%%%%%%%%%%%%%%%%%%%%%%%%%%%%%%
Then, the second \cred{zero-energy} solution orthogonal to $\Phi^{(1)}_{E=0-}$ is found to be 
	%%%%%%%%%%%%%%%%%%%%%%%%%%%%%%%%%%%%%%%%%%%%%%%%%%%%%%%%%%%%%%%%%%%%%%%%%%
	\begin{align}
	\Phi^{(2)}_{E=0-}(y)=\left(\begin{array}{c}
						0\\
						0\\
						N'e^{-i\frac{\theta}{2}}e^{-M(\cos\theta\,y_{1}+\sin\theta\,y_{2})}\\
						-N'e^{i\frac{\theta}{2}}e^{-M(\cos\theta\,y_{1}+\sin\theta\,y_{2})}
						\end{array}\right)\label{zero-mode-solution-2-}
	\end{align}
	%%%%%%%%%%%%%%%%%%%%%%%%%%%%%%%%%%%%%%%%%%%%%%%%%%%%%%%%%%%%%%%%%%%%%%%%%%
with an appropriate normalization constant $N'$.

Since there \cred{are} no more independent \cred{zero-energy solutions} of the type (\ref{another-type-of-zero-energy-solution}), we may conclude that the number of the degeneracy of the \cred{zero-energy} solutions is two.
This result seems to be consistent with the degeneracy of the \cred{positive-energy} solutions $\Phi^{(i)}_{E_{n_{1}n_{2}-}}$ with $i=1,2$.

Before closing this subsection, we would like to \cred{comment} on a general form of \cred{zero-energy} solutions.
We first note that the wavefunction (\ref{zero-mode-solution-1-}) satisfies the \cred{zero-energy} equation (\ref{zero-energy-equations-g1g2}) even for an arbitrary complex number $\theta$. Then, we can show that a general form of \cred{zero-energy} solutions to (\ref{zero-energy-equations-g1g2}) is given by the superposition of the solution (\ref{zero-mode-solution-1-}) with respect to $\theta \in \mathbb{C}$. It follows from this observation that additional conditions (for instance, additional boundary conditions like $\partial_{y_{2}}g_{1}(y)=\partial_{y_{2}}g_{2}(y)=0$ at $y_{1}=0,L_{1}$ and $y_{2}=0,L_{2}$) seem to be necessary to determine independent \cred{zero-energy} solutions definitely.

%%%%%%%%%%%%%%%		Subsection~5.4		%%%%%%%%%%%%%%%%%%%
\subsection{\cred{Four-dimensional} mass spectrum}
\label{sec:4d mass spectrum}

In the previous subsections, we have succeeded in obtaining the energy spectrum of the Hamiltonian system $H=Q^{2}$, though we have not yet arrived at a definite conclusion for \cred{zero-energy} solutions. We can use those results to expand the original 6d Dirac field $\Psi (x,y)$ in the 4d \cred{Kaluza--Klein} modes, and then rewrite the action (\ref{Dirac-action}) into the \cred{four}-dimensional effective action that consists of an infinite number of 4d massive fermions and a finite number of 4d massless chiral ones.

As discussed in Section \ref{sec:6d Dirac fermion on a rectangle}, the 6d Dirac field $\Psi (x,y)$ can be decomposed into the eigenfunctions of $\Gamma^{5}$ and $\Gamma^{y}$ as
	%%%%%%%%%%%%%%%%%%%%%%%%%%%%%%%%%%%%%%%%%%%%%%%%%%%%%%%%%%%%%%%%%%%%%%%%%%
	\begin{align}
	\Psi(x,y)=\Psi_{R+}(x,y)+\Psi_{R-}(x,y)+\Psi_{L+}(x,y)+\Psi_{L-}(x,y),\label{RL+--decomposition}
	\end{align}
	%%%%%%%%%%%%%%%%%%%%%%%%%%%%%%%%%%%%%%%%%%%%%%%%%%%%%%%%%%%%%%%%%%%%%%%%%%
where the subscripts $\pm$ of $\Psi_{R\pm}$ and $\Psi_{L\pm}$ denote the eigenvalues of $\Gamma^{y}$ (but not $(-1)^{F}$).\footnote{We hope that \cred{readers} do not confuse the meanings of the subscripts $\pm$ for $\Psi_{R\pm}$, $\Psi_{L\pm}$ in (\ref{RL+--decomposition}) with $\Phi_{E\pm}(y)$ in (\ref{-1F-eigenstates}).}
The results given in the previous subsections suggest that {$\Psi_{R\pm}(x,y)$ and $\Psi_{L\pm}(x,y)$}  may be expanded, in terms of the energy eigenfunctions, as 
	%%%%%%%%%%%%%%%%%%%%%%%%%%%%%%%%%%%%%%%%%%%%%%%%%%%%%%%%%%%%%%%%%%%%%%%%%%
	\begin{align}
	&\Psi_{R\pm}(x,y)=\sum^{\infty}_{n_{1}=1}\sum^{\infty}_{n_{2}=1}\psi^{(n_{1},n_{2})}_{R\pm}(x)f_{n_{1}n_{2}}(y),\nonumber\\
	&\Psi_{L\pm}(x,y)=\Psi^{(0)}_{L\pm}(x,y)\nonumber\\
	&{+\sum^{\infty}_{n_{1}=1}\sum^{\infty}_{n_{2}=1}\left\{ i\Gamma^{y_{1}}\eta_{L\pm}^{(n_{1},n_{2})}(x)\frac{1}{\sqrt{E_{n_{1}n_{2}}}}(\partial_{y_{1}} {\mp} i\partial_{y_{2}})f_{n_{1}n_{2}}(y)+\frac{M}{\sqrt{E_{n_{1}n_{2}}}}\eta^{(n_{1},n_{2})}_{L\pm}(x)f_{n_{1}n_{2}}(y)\right\}}{,}\label{mode-expansion-Lpm}
	\end{align}
	%%%%%%%%%%%%%%%%%%%%%%%%%%%%%%%%%%%%%%%%%%%%%%%%%%%%%%%%%%%%%%%%%%%%%%%%%%
where
	%%%%%%%%%%%%%%%%%%%%%%%%%%%%%%%%%%%%%%%%%%%%%%%%%%%%%%%%%%%%%%%%%%%%%%%%%%
	\begin{align}
	&\Psi^{(0)}_{L+}(x,y)=\xi^{(0)}_{L1}(x){N}e^{-i\frac{\theta}{2}}e^{M(\cos\theta\,y_{1}+\sin\theta\, y_{2})}+\xi^{(0)}_{L2}(x){N'}e^{-i\frac{\theta}{2}}e^{-M(\cos\theta\,y_{1}+\sin\theta\, y_{2})},\nonumber\\
	&\Psi^{(0)}_{L-}(x,y)=i\Gamma^{y_{1}}\xi^{(0)}_{L1}(x){N}e^{i\frac{\theta}{2}}e^{M(\cos\theta\,y_{1}+\sin\theta\, y_{2})}-i\Gamma^{y_{1}}\xi^{(0)}_{L2}(x){N'}e^{i\frac{\theta}{2}}e^{-M(\cos\theta\,y_{1}+\sin\theta\, y_{2})}.\label{mode-expantion-zero-energy}
	\end{align}
	%%%%%%%%%%%%%%%%%%%%%%%%%%%%%%%%%%%%%%%%%%%%%%%%%%%%%%%%%%%%%%%%%%%%%%%%%%
Here, $\psi^{(n_{1},n_{2})}_{R\pm}(x)$, $\eta_{L\pm}^{(n_{1},n_{2})}(x)$ and $\xi^{(0)}_{Li}(x)$ ($i=1,2$) denote 4d chiral spinors as depicted by the subscripts $R$ and $L$. We would like to note that the form of the mode expansion of $\Psi_{L\pm} (x,y)$ is not trivial and that the mode expansions of $\Psi_{R\pm}(x,y)$ and $\Psi_{L\pm}(x,y)$ have to be arranged such that $\psi^{(n_{1},n_{2})}_{R\pm}(x)$, $\eta_{L\pm}^{(n_{1},n_{2})}(x)$ and $\xi^{(0)}_{Li}(x)$ give the 4d mass eigenstates.

By inserting the expansions (\ref{mode-expansion-Lpm}) and (\ref{mode-expantion-zero-energy}) into the original action (\ref{Dirac-action}) and integrating over $y_{1}$ and $y_{2}$, we find that the action (\ref{Dirac-action}) becomes\,\footnote{The results are consistent with those given in Ref.~{\cite{Fujimoto:2016llj}}.}
	%%%%%%%%%%%%%%%%%%%%%%%%%%%%%%%%%%%%%%%%%
	%%%%%%%%%%%%%%%%%%%%%%%%%%%%%%%%%%%%%%%%%%%%%%%%%%%%%%%%%%%%%%%%%%%%%%%%%%
	\begin{align}
	&S=\int d^{4}x \left\{\sum^{2}_{i=1}\overline{\xi}^{(0)}_{Li}(x)\,i\Gamma^{\mu}\partial_{\mu}\xi^{(0)}_{Li}(x)\right.\nonumber\\
	&\left.+\sum^{\infty}_{n_{1}=1}\sum^{\infty}_{n_{2}=1}\biggl[\overline{\psi}_{1}^{(n_{1},n_{2})}(x)\Bigl(i\Gamma^{\mu}\partial_{\mu}-m_{n_{1},n_{2}}\Bigr)\psi_{1}^{(n_{1},n_{2})}(x)+\overline{\psi}^{(n_{1},n_{2})}_{2}(x)\Bigl(i\Gamma^{\mu}\partial_{\mu}-m_{n_{1},n_{2}}\Bigr)\psi^{(n_{1},n_{2})}_{2}(x)\biggr]\right\},
	\end{align}
	%%%%%%%%%%%%%%%%%%%%%%%%%%%%%%%%%%%%%%%%%%%%%%%%%%%%%%%%%%%%%%%%%%%%%%%%%%
where $\psi^{(n_{1},n_{2})}_{i}(x)$ are 4d Dirac spinors defined by
	%%%%%%%%%%%%%%%%%%%%%%%%%%%%%%%%%%%%%%%%%%%%%%%%%%%%%%%%%%%%%%%%%%%%%%%%%%
	\begin{align}
	\psi_{1}^{(n_{1},n_{2})}(x)\equiv \psi^{(n_{1},n_{2})}_{R+}(x)+\eta^{(n_{1},n_{2})}_{L-}(x),\nonumber\\
	\psi_{2}^{(n_{1},n_{2})}(x)\equiv \psi^{(n_{1},n_{2})}_{R-}(x)+\eta^{(n_{1},n_{2})}_{L+}(x),
	\end{align}
	%%%%%%%%%%%%%%%%%%%%%%%%%%%%%%%%%%%%%%%%%%%%%%%%%%%%%%%%%%%%%%%%%%%%%%%%%%
and
	%%%%%%%%%%%%%%%%%%%%%%%%%%%%%%%%%%%%%%%%%%%%%%%%%%%%%%%%%%%%%%%%%%%%%%%%%%
	\begin{align}
	m_{n_{1}n_{2}}\equiv \sqrt{E_{n_{1}n_{2}}}=\sqrt{\left(\frac{n_{1}\pi}{L_{1}}\right)^{2}+\left(\frac{n_{2}\pi}{L_{2}}\right)^{2}+M^{2}}.
	\end{align}
	%%%%%%%%%%%%%%%%%%%%%%%%%%%%%%%%%%%%%%%%%%%%%%%%%%%%%%%%%%%%%%%%%%%%%%%%%%
Thus, we conclude that the 4d mass spectrum of the 6d Dirac fermion for \cred{the} {Type \GkII} boundary condition consists of infinitely many massive Dirac fermions $\psi_{i}^{(n_{1},n_{2})}(x)$ ($n_{1}, n_{2} =1,2,3,\cred{\ldots};\ i=1,2$) with mass $m_{n_{1}n_{2}}$ and two massless left-handed chiral fermions $\xi^{(0)}_{Li}(x)$ ($i=1,2$). It should be emphasized that the appearance of the degenerate massless chiral fermions in the 4d mass spectrum could have important implications for phenomenology to solve the generation problem of the quarks and leptons.

%%%%%%%%%%%%%%%%%%%%%%%%%%%%%%%%%%%%%
%%%%%%%% Section 6 %%%%%%%%%%%%%%%%%%
%%%%%%%%%%%%%%%%%%%%%%%%%%%%%%%%%%%%%
\section{Energy spectrum for {Type \GkIII} \cred{boundary conditions}
\label{sec:Energy spectrum for type III BC}}

In this section, we investigate the energy spectrum for {Type \GkIII} boundary condition in a slightly different way \cred{than} in the previous section.

%%%%%%%%%%%%%%%		Subsection~6.1		%%%%%%%%%%%%%%%%%%%
\subsection{Type {\GkIII} \cred{boundary conditions} and reformulation of SUSY}
\label{sec:Type III BC and reformulation of SUSY}

{Type \GkIII} boundary condition has the $S^{2}$ \cred{parameters} at each boundary of $y_{1}=0,L_{1}$ and $y_{2}=0,L_{2}$. {For simplicity in the following, we restrict our considerations to \cred{the} simple case of }
	%%%%%%%%%%%%%%%%%%%%%%%%%%%%%%%%%%%%%%%%%%%%%%%%%%%%%%%%%%%%%%%%%%%%%%%%%%
	\begin{align}
	\vec{n}_{1}=\vec{n}_{2}=(0,0,-1)\equiv \vec{n}
	\end{align}
	%%%%%%%%%%%%%%%%%%%%%%%%%%%%%%%%%%%%%%%%%%%%%%%%%%%%%%%%%%%%%%%%%%%%%%%%%%
for the $S^{2}$ \cred{parameters}. Then, the boundary condition considered in this section is given by
	%%%%%%%%%%%%%%%%%%%%%%%%%%%%%%%%%%%%%%%%%%%%%%%%%%%%%%%%%%%%%%%%%%%%%%%%%%
	\begin{align}
	\Phi_{\vec{n}\cdot\vec{\Sigma}_{1}=-1}(y)=\Phi_{\vec{n}\cdot\vec{\Sigma}_{2}=-1}(y)=\left(\begin{array}{c}
																		f_{1}(y)\\
																		0\\
																		g_{1}(y)\\
																		0
																		\end{array}\right)=0\qquad \text{at}\ \ y_{1}=0,L_{1}\ \ \text{and}\ \ y_{2}=0,L_{2}. \label{typeIII-BC-simple-choice}
	\end{align}
	%%%%%%%%%%%%%%%%%%%%%%%%%%%%%%%%%%%%%%%%%%%%%%%%%%%%%%%%%%%%%%%%%%%%%%%%%%

Although we could follow the previous analysis for the \cred{Type} {\GkII} boundary condition, it will be convenient to reformulate the Hamiltonian with a different supercharge. By decomposing $\Psi(x,y)$ into the eigenstates of $\Gamma^{y}$ as $\Psi(x,y)=\Psi_{+}(x,y)+\Psi_{-}(x,y)$, we may rewrite the Dirac equation (\ref{Dirac-equation}) into the form
	%%%%%%%%%%%%%%%%%%%%%%%%%%%%%%%%%%%%%%%%%%%%%%%%%%%%%%%%%%%%%%%%%%%%%%%%%%
	\begin{align}
	\left(\begin{array}{cc}
		i\Gamma^{\mu}\partial_{\mu}-M&0\\
		0&i\Gamma^{\mu}\partial_{\mu}+M
		\end{array}\right)\left(\begin{array}{c}
						\Psi_{+}(x,y)\\
						\widetilde{\Psi}_{+}(x,y)
						\end{array}\right) =\left(\begin{array}{cc}
											0&-(\partial_{y_{1}}-i\partial_{y_{2}})\\
											\partial_{y_{1}}+i\partial_{y_{2}}&0
											\end{array}\right)\left(\begin{array}{c}
																\Psi_{+}(x,y)\\
																\widetilde{\Psi}_{+}(x,y)
																\end{array}\right),
	\end{align}
	%%%%%%%%%%%%%%%%%%%%%%%%%%%%%%%%%%%%%%%%%%%%%%%%%%%%%%%%%%%%%%%%%%%%%%%%%%
where $\widetilde{\Psi}_{+}(x,y)\equiv i\Gamma^{y_{1}}\Psi_{-}(x,y)$.

We can then define a new Hamiltonian $\widetilde{H}$ by
	%%%%%%%%%%%%%%%%%%%%%%%%%%%%%%%%%%%%%%%%%%%%%%%%%%%%%%%%%%%%%%%%%%%%%%%%%%
	\begin{align}
	\widetilde{H}\equiv \widetilde{Q}^{2}=\Bigl[-(\partial_{y_{1}})^{2}-(\partial_{y_{2}})^{2}\Bigr]{\rm I}_{2},
	\end{align}
	%%%%%%%%%%%%%%%%%%%%%%%%%%%%%%%%%%%%%%%%%%%%%%%%%%%%%%%%%%%%%%%%%%%%%%%%%%
with a new supercharge
	%%%%%%%%%%%%%%%%%%%%%%%%%%%%%%%%%%%%%%%%%%%%%%%%%%%%%%%%%%%%%%%%%%%%%%%%%%
	\begin{align}
	\widetilde{Q}\equiv \left(\begin{array}{cc}
											0&-(\partial_{y_{1}}-i\partial_{y_{2}})\\
											\partial_{y_{1}}+i\partial_{y_{2}}&0
											\end{array}\right).
	\end{align}
	%%%%%%%%%%%%%%%%%%%%%%%%%%%%%%%%%%%%%%%%%%%%%%%%%%%%%%%%%%%%%%%%%%%%%%%%%%
Here, $\widetilde{H}$ and $\widetilde{Q}$ are represented by $2\times 2$ matrices, instead of $4\times 4$. The differential operators $\widetilde{H}$ and $\widetilde{Q}$ act on the two-component wavefunction
	%%%%%%%%%%%%%%%%%%%%%%%%%%%%%%%%%%%%%%%%%%%%%%%%%%%%%%%%%%%%%%%%%%%%%%%%%%
	\begin{align}
	\widetilde{\Phi}(y)=\left(\begin{array}{c}
						\widetilde{f}(y)\\
						\widetilde{g}(y)
						\end{array}\right)
	\end{align}
	%%%%%%%%%%%%%%%%%%%%%%%%%%%%%%%%%%%%%%%%%%%%%%%%%%%%%%%%%%%%%%%%%%%%%%%%%%
with the boundary condition
	%%%%%%%%%%%%%%%%%%%%%%%%%%%%%%%%%%%%%%%%%%%%%%%%%%%%%%%%%%%%%%%%%%%%%%%%%%
	\begin{align}
	\widetilde{f}(y)=0\qquad \text{at}\ \ y_{1}=0,L_{1}\ \ \text{and}\ \ y_{2}=0,L_{2} \label{typeIII-BC-rewritten}
	\end{align}
	%%%%%%%%%%%%%%%%%%%%%%%%%%%%%%%%%%%%%%%%%%%%%%%%%%%%%%%%%%%%%%%%%%%%%%%%%%
which will correspond to (\ref{typeIII-BC-simple-choice}). It should be stressed that the above boundary condition (\ref{typeIII-BC-rewritten}) guarantees \cred{that the supercharge $\widetilde{Q}$ is Hermitian}.

The ``fermion'' number operator $\widetilde{F}$ can be introduced as 
	%%%%%%%%%%%%%%%%%%%%%%%%%%%%%%%%%%%%%%%%%%%%%%%%%%%%%%%%%%%%%%%%%%%%%%%%%%
	\begin{align}
	(-1)^{\widetilde{F}}=\left(\begin{array}{cc}
						1&0\\
						0&-1
						\end{array}\right)
	\end{align}
	%%%%%%%%%%%%%%%%%%%%%%%%%%%%%%%%%%%%%%%%%%%%%%%%%%%%%%%%%%%%%%%%%%%%%%%%%%
which satisfies all the desired relations discussed in the previous sections.

%%%%%%%%%%%%%%%		Subsection~6.2		%%%%%%%%%%%%%%%%%%%
\subsection{Energy spectrum}
\label{sec:Energy spectrum}

In order to construct the energy spectrum, it will be convenient to introduce the eigenfunctions of $(-1)^{\widetilde{F}}$, such that
	%%%%%%%%%%%%%%%%%%%%%%%%%%%%%%%%%%%%%%%%%%%%%%%%%%%%%%%%%%%%%%%%%%%%%%%%%%
	\begin{align}
	(-1)^{\widetilde{F}}\widetilde{\Phi}_{\pm}(y)=\pm\widetilde{\Phi}_{\pm}(y),
	\end{align}
	%%%%%%%%%%%%%%%%%%%%%%%%%%%%%%%%%%%%%%%%%%%%%%%%%%%%%%%%%%%%%%%%%%%%%%%%
where
	%%%%%%%%%%%%%%%%%%%%%%%%%%%%%%%%%%%%%%%%%%%%%%%%%%%%%%%%%%%%%%%%%%%%%%%%%%
	\begin{align}
	\widetilde{\Phi}_{+}(y)=\left(\begin{array}{c}
							\widetilde{f}(y)\\
							0
							\end{array}\right),\qquad \widetilde{\Phi}_{-}(y)=\left(\begin{array}{c}
																	0\\
																	\widetilde{g}(y)
																	\end{array}\right).
	\end{align}
	%%%%%%%%%%%%%%%%%%%%%%%%%%%%%%%%%%%%%%%%%%%%%%%%%%%%%%%%%%%%%%%%%%%%%%%%%%

With the boundary conditions (\ref{typeIII-BC-rewritten}), we can easily find the energy eigenfunctions for $\widetilde{\Phi}_{+}(y)$. The result is  
	%%%%%%%%%%%%%%%%%%%%%%%%%%%%%%%%%%%%%%%%%%%%%%%%%%%%%%%%%%%%%%%%%%%%%%%%%%
	\begin{align}
	&\widetilde{H}\widetilde{\Phi}_{\widetilde{E}_{n_{1}n_{2}}+}(y)=\widetilde{E}_{n_{1}n_{2}}\widetilde{\Phi}_{\widetilde{E}_{n_{1}n_{2}}+}(y),\nonumber\\
	&\widetilde{\Phi}_{\widetilde{E}_{n_{1}n_{2}}+}(y)=\left(\begin{array}{c}	
											f_{n_{1}n_{2}}(y)\\
											0
											\end{array}\right) \qquad (n_{1},n_{2}=1,2,3,\cred{\ldots}){,}
	\end{align}
	%%%%%%%%%%%%%%%%%%%%%%%%%%%%%%%%%%%%%%%%%%%%%%%%%%%%%%%%%%%%%%%%%%%%%%%%%%
where $f_{n_{1}n_{2}}(y)$ are defined in (\ref{fn1n2}) and 
	%%%%%%%%%%%%%%%%%%%%%%%%%%%%%%%%%%%%%%%%%%%%%%%%%%%%%%%%%%%%%%%%%%%%%%%%%%
	\begin{align}
	\widetilde{E}_{n_{1}n_{2}}=\left(\frac{n_{1}\pi}{L_{1}}\right)^{2}+{\left(\frac{n_{2}\pi}{L_{2}}\right)^{2}}\,\qquad (n_{1},n_{2}=1,2,3,\cred{\ldots}).
	\end{align}
	%%%%%%%%%%%%%%%%%%%%%%%%%%%%%%%%%%%%%%%%%%%%%%%%%%%%%%%%%%%%%%%%%%%%%%%%%%
	
In order to obtain the \cred{positive-energy} spectrum for $\widetilde{\Phi}_{-}(y)$, we use the SUSY relations
	%%%%%%%%%%%%%%%%%%%%%%%%%%%%%%%%%%%%%%%%%%%%%%%%%%%%%%%%%%%%%%%%%%%%%%%%%%
	\begin{align}
	\sqrt{\widetilde{E}_{n_{1}n_{2}}}\widetilde{\Phi}_{\widetilde{E}_{n_{1}n_{2}}\mp}(y)=\widetilde{Q}\widetilde{\Phi}_{\widetilde{E}_{n_{1}n_{2}}\pm}(y).\label{typeIII-BC-SUSY-relation}
	\end{align}
	%%%%%%%%%%%%%%%%%%%%%%%%%%%%%%%%%%%%%%%%%%%%%%%%%%%%%%%%%%%%%%%%%%%%%%%%%%
It follows that the \cred{positive-energy} eigenfunctions $\widetilde{\Phi}_{\widetilde{E}_{n_{1}n_{2}}-}(y)$ are given by
	%%%%%%%%%%%%%%%%%%%%%%%%%%%%%%%%%%%%%%%%%%%%%%%%%%%%%%%%%%%%%%%%%%%%%%%%%%
	\begin{align}
	\widetilde{\Phi}_{\widetilde{E}_{n_{1}n_{2}}-}(y)=\left(\begin{array}{c}
											0\\
											\frac{1}{\sqrt{\widetilde{E}_{n_{1}n_{2}}}}(\partial_{y_{1}}+i\partial_{y_{2}})f_{n_{1}n_{2}}(y)
											\end{array}\right).
	\end{align}
	%%%%%%%%%%%%%%%%%%%%%%%%%%%%%%%%%%%%%%%%%%%%%%%%%%%%%%%%%%%%%%%%%%%%%%%%%%
The SUSY relations (\ref{typeIII-BC-SUSY-relation}) also imply that $\widetilde{\Phi}_{-}(y)$ should satisfy the \cred{boundary condition}
	%%%%%%%%%%%%%%%%%%%%%%%%%%%%%%%%%%%%%%%%%%%%%%%%%%%%%%%%%%%%%%%%%%%%%%%%%%
	\begin{align}
	\widetilde{Q}\widetilde{\Phi}_{-}(y)=0\qquad \text{at}\ \ y_{1}=0,L_{1}\ \ \text{and}\ \ y_{2}=0,L_{2},\nonumber
	\end{align}
	%%%%%%%%%%%%%%%%%%%%%%%%%%%%%%%%%%%%%%%%%%%%%%%%%%%%%%%%%%%%%%%%%%%%%%%%%%
or equivalently
	%%%%%%%%%%%%%%%%%%%%%%%%%%%%%%%%%%%%%%%%%%%%%%%%%%%%%%%%%%%%%%%%%%%%%%%%%%
	\begin{align}
	(\partial_{y_{1}}-i\partial_{y_{2}})\widetilde{g}(y)=0\qquad \text{at}\ \ y_{1}=0,L_{1}\ \ \text{and}\ \ y_{2}=0,L_{2}.\label{typeIII-BC-tildeg}
	\end{align}
	%%%%%%%%%%%%%%%%%%%%%%%%%%%%%%%%%%%%%%%%%%%%%%%%%%%%%%%%%%%%%%%%%%%%%%%%%%

This is not the end of the story. The set of $\{\widetilde{\Phi}_{\widetilde{E}_{n_{1}n_{2}}\pm}(y);\ n_{1},n_{2}=1,2,3,\cred{\ldots}\}$ gives a complete spectrum for the \cred{positive-energy} state, but we have not yet obtained \cred{zero-energy} eigenfunctions for $\widetilde{\Phi}_{E=0-}(y)$.

Since $\widetilde{\Phi}_{+}(y)$ obeys the Dirichlet boundary condition, it cannot possess any \cred{zero-energy} state. Therefore, any \cred{zero-energy} solution to $\widetilde{H}=\widetilde{Q}^{2}$ should appear from an eigenstate of $(-1)^{\widetilde{F}}=-1$ and satisfies $\widetilde{Q}\widetilde{\Phi}_{E=0-}(y)=0$, i.e.
	%%%%%%%%%%%%%%%%%%%%%%%%%%%%%%%%%%%%%%%%%%%%%%%%%%%%%%%%%%%%%%%%%%%%%%%%%%
	\begin{align}
	(\partial_{y_{1}}-i\partial_{y_{2}})\widetilde{g}_{E=0}(y)=0. \label{typeIII-BC-zeromodeeq--}
	\end{align}
	%%%%%%%%%%%%%%%%%%%%%%%%%%%%%%%%%%%%%%%%%%%%%%%%%%%%%%%%%%%%%%%%%%%%%%%%%%
A general solution to (\ref{typeIII-BC-zeromodeeq--}) is given by
	%%%%%%%%%%%%%%%%%%%%%%%%%%%%%%%%%%%%%%%%%%%%%%%%%%%%%%%%%%%%%%%%%%%%%%%%%%
	\begin{align}
	{\widetilde{g}_{E=0}(y)=\rho(\overline{z}),}\label{typeIII-BC-zero-energy-solution}
	\end{align}
	%%%%%%%%%%%%%%%%%%%%%%%%%%%%%%%%%%%%%%%%%%%%%%%%%%%%%%%%%%%%%%%%%%%%%%%%%%
where {$\rho (\overline{z})$} is an arbitrary {anti-}holomorphic function of {$\overline{z}=y_{1}-iy_{2}$}.

Here, we face a strange situation again. The \cred{Type} I\!I\!I boundary condition (\ref{typeIII-BC-rewritten}) for $\widetilde{\Phi}_{+}(y)$ and (\ref{typeIII-BC-tildeg}) for $\widetilde{\Phi}_{-}(y)$ \cred{turns} out to work well to determine the \cred{positive-energy} solutions. On the other hand, the boundary condition (\ref{typeIII-BC-tildeg}) for $\widetilde{\Phi}_{-}(y)$ or $\widetilde{g}(y)$ does not work properly for \cred{zero-energy} solutions because any \cred{zero-energy} solution to (\ref{typeIII-BC-zeromodeeq--}) trivially satisfies the boundary condition (\ref{typeIII-BC-tildeg}), and in fact the boundary condition does not give any restriction on \cred{zero-energy} solutions.

It \cred{is worth commenting} on a general form of \cred{zero-energy} solutions (\ref{typeIII-BC-zero-energy-solution}).
The \cred{zero-energy} equation $\widetilde{Q}\widetilde{\Phi}_{E=0}(y)=0$ possesses \cred{two}-dimensional conformal invariance because $\tilde{Q}$ includes no massive parameter. Therefore, it is reasonable that a general solution to the conformal invariant equation $\widetilde{Q}\widetilde{\Phi}(y)=0$ is given by any {anti-holomorphic} function (without specifying non-trivial boundary conditions).

%%%%%%%%%%%%%%%%%%%%%%%%%%%%%%%%%%%%%
%%%%%%%% Section 7 %%%%%%%%%%%%%%%%%%
%%%%%%%%%%%%%%%%%%%%%%%%%%%%%%%%%%%%%
\section{Mapping between degenerate states
\label{sec:Mapping between degenerate states}}

In Section \ref{sec:Energy spectrum for type II BC}, we have found that \cred{positive-energy} eigenfunctions are four-fold degenerate for \cred{the} {Type \GkII} boundary condition. The purpose of this section is to understand the degeneracy of the energy eigenfunctions, especially for the \cred{positive-energy} states. In the following analysis, we will restrict our considerations to the energy spectrum for \cred{the} {Type \GkII} boundary condition.

As already discussed, every pair of \cred{positive-energy} eigenfunctions $\Phi_{E+}$ and $\Phi_{E-}$ forms a supermultiplet. This implies that the \cred{positive-energy} solutions $\Phi^{(i)}_{E_{n_{1}n_{2}}+}$ ($n_{1},n_{2}=1,2,3,\cred{\ldots};\ i=1,2$) are related to $\Phi^{(i)}_{E_{n_{1}n_{2}}-}$ by supersymmetry, i.e.
	%%%%%%%%%%%%%%%%%%%%%%%%%%%%%%%%%%%%%%%%%%%%%%%%%%%%%%%%%%%%%%%%%%%%%%%%%%
	\begin{align}
	\Phi^{(1)}_{E_{n_{1}n_{2}}+}\xlongleftrightarrow{\text{\scriptsize{Q}}} \Phi^{(1)}_{E_{n_{1}n_{2}}-}{,}\nonumber\\
	\Phi^{(2)}_{E_{n_{1}n_{2}}+}\xlongleftrightarrow{\text{\scriptsize{Q}}} \Phi^{(2)}_{E_{n_{1}n_{2}}-}{.}
	%\overset{Q}{\overleftrightarrow}
	\end{align}
	%%%%%%%%%%%%%%%%%%%%%%%%%%%%%%%%%%%%%%%%%%%%%%%%%%%%%%%%%%%%%%%%%%%%%%%%%%

To clarify the relations between $\Phi^{(1)}_{E_{n_{1}n_{2}}\pm}$ and $\Phi^{(2)}_{E_{n_{1}n_{2}}\pm}$, let us consider the ${\cal C}$ \cred{transformation} defined by\hspace{-0.2em}
	%%%%%%%%%%%%%%%%%%%%%%%%%%%%%%%%%%%%%%%%%%%%%%%%%%%%%%%%%%%%%%%%%%%%%%%%%%
	\begin{align}
	\Phi(y)\xlongrightarrow{{\scriptsize{\cal C}}}{\cal C}\Phi(y)\equiv C\Bigl(\Phi(y)\Bigr)^{\ast},\label{Ctransformation}
	\end{align}
	%%%%%%%%%%%%%%%%%%%%%%%%%%%%%%%%%%%%%%%%%%%%%%%%%%%%%%%%%%%%%%%%%%%%%%%%%%
where $C$ is the $4\times 4$ matrix
	%%%%%%%%%%%%%%%%%%%%%%%%%%%%%%%%%%%%%%%%%%%%%%%%%%%%%%%%%%%%%%%%%%%%%%%%%%
	\begin{align}
	C\equiv \left(\begin{array}{cc}
					\sigma_{1}&0\\
					0&-\sigma_{1}
					\end{array}\right).
	\end{align}
	%%%%%%%%%%%%%%%%%%%%%%%%%%%%%%%%%%%%%%%%%%%%%%%%%%%%%%%%%%%%%%%%%%%%%%%%%%
Interestingly, we can show that the ${\cal C}$ \cred{transformation} satisfies the following relations,
	%%%%%%%%%%%%%%%%%%%%%%%%%%%%%%%%%%%%%%%%%%%%%%%%%%%%%%%%%%%%%%%%%%%%%%%%%%
	\begin{align}
	{\cal C}(-1)^{F}&=(-1)^{F}{\cal C},\nonumber\\
	{\cal C}Q&=Q{\cal C},\nonumber\\
	{\cal C}H&=H{\cal C},\nonumber\\
	({\cal C})^{2}&=1. \label{C-transformaion-property}
	\end{align}
	%%%%%%%%%%%%%%%%%%%%%%%%%%%%%%%%%%%%%%%%%%%%%%%%%%%%%%%%%%%%%%%%%%%%%%%%%%

It follows from (\ref{C-transformaion-property}) that if $\Phi_{E\pm}(y)$ are any eigenfunctions of $H=E$ and $(-1)^{F}=\pm 1$, then the states ${\cal C}\Phi_{E\pm}(y)$ also have the same eigenvalues as $\Phi_{E\pm}(y)$, i.e. 
	%%%%%%%%%%%%%%%%%%%%%%%%%%%%%%%%%%%%%%%%%%%%%%%%%%%%%%%%%%%%%%%%%%%%%%%%%%
	\begin{align}
	H\Bigl({\cal C}\Phi_{E\pm}(y)\Bigr)=E\Bigl({\cal C}\Phi_{E\pm}(y)\Bigr),\\
	(-1)^{F}\Bigl({\cal C}\Phi_{E\pm}(y)\Bigr)=\pm\Bigl({\cal C}\Phi_{E\pm}(y)\Bigr).
	\end{align}
	%%%%%%%%%%%%%%%%%%%%%%%%%%%%%%%%%%%%%%%%%%%%%%%%%%%%%%%%%%%%%%%%%%%%%%%%%%
If ${\cal C}\Phi_{E\pm}$ are not proportional to $\Phi_{E\pm}$ themselves, $\Phi_{E\pm}(y)$ and ${\cal C}\Phi_{E\pm}(y)$ can be independent \cred{of} each other with the same energy eigenvalue $E$. This observation implies that the set of $\{\Phi_{E\pm},{\cal C}\Phi_{E\pm}\}$ gives four-fold degenerate eigenstates of $H=E$. In fact, the eigenfunctions $\{\Phi^{(1)}_{E_{n_{1}n_{2}}\pm},\Phi^{(2)}_{E_{n_{1}n_{2}}\pm}\}$ turn out to be related as
	%%%%%%%%%%%%%%%%%%%%%%%%%%%%%%%%%%%%%%%%%%%%%%%%%%%%%%%%%%%%%%%%%%%%%%%%%%
	\begin{align}
		\begin{array}{ccc}
		\Phi^{(1)}_{E_{n_{1}n_{2}}+}&\xlongleftrightarrow{Q}&\Phi^{(1)}_{E_{n_{1}n_{2}}-}\\[0.5cm]
		\biggl\updownarrow\ {\cal C}&&\biggr\updownarrow\ {\cal C}\\[0.5cm]
		\Phi^{(2)}_{E_{n_{1}n_{2}}+}&\xlongleftrightarrow{Q}&\Phi^{(2)}_{E_{n_{1}n_{2}}-}
		\end{array}{.}
	\end{align}
	%%%%%%%%%%%%%%%%%%%%%%%%%%%%%%%%%%%%%%%%%%%%%%%%%%%%%%%%%%%%%%%%%%%%%%%%%%
For the \cred{zero-energy} eigenfunctions $\Phi^{(1)}_{E=0-}$ and $\Phi^{(2)}_{E=0-}$ given in (\ref{zero-mode-solution-1-}) and (\ref{zero-mode-solution-2-}), we find\\[0.2cm]
	%%%%%%%%%%%%%%%%%%%%%%%%%%%%%%%%%%%%%%%%%%%%%%%%%%%%%%%%%%%%%%%%%%%%%%%%%%
	\begin{align}
	{\cal C} \ \rotatebox[origin=c]{-90}{\LARGE$\circlearrowleft$}\ \Phi^{(1)}_{E=0-}\xlongrightarrow{\LARGE Q}0\xlongleftarrow{Q}\Phi^{(2)}_{E=0-}\ \rotatebox[origin=c]{90}{\LARGE$\circlearrowright$}\ {\cal C}{,}
	\end{align}
	%%%%%%%%%%%%%%%%%%%%%%%%%%%%%%%%%%%%%%%%%%%%%%%%%%%%%%%%%%%%%%%%%%%%%%%%%%
where $\Phi^{(1)}_{E=0-}$ and $\Phi^{(2)}_{E=0-}$ are found to be eigenfunctions of ${\cal C}=-1$ and ${\cal C}=+1$, respectively.

{In the following part, we show that this ${\cal C}$ transformation for mode functions originates from a CP transformation in a 6d sense}.
{Let us consider} a CP transformation that consists of the 6d charge conjugation $C$ and parity transformation $P$ with $(t,{\bm x},y)\rightarrow\ (t,-{\bm x},y)$. The 6d charge conjugation is given by
	%%%%%%%%%%%%%%%%%%%%%%%%%%%%%%%%%%%%%%%%%%%%%%%%%%%%%%%%%%%%%%
	\begin{align}
	C:\Psi(x,y)\rightarrow\ \Psi^{(C)}(x,y)=C\overline{\Psi}^{\cred{\rm T}}(x,y),
	\end{align}
	%%%%%%%%%%%%%%%%%%%%%%%%%%%%%%%%%%%%%%%%%%%%%%%%%%%%%%%%%%%%%%
where $C$ is {an} $8\times8$ unitary matrix.
{The} concrete definition and properties of the 6d charge conjugation are given in \cred{Appendix}~\ref{sec:appendixB}. 
This transformation flips both \cred{the} 4d chirality $R/L$ and the inner chirality $\pm$ {(see \cred{Appendix}~\ref{sec:appendixA})} as
	%%%%%%%%%%%%%%%%%%%%%%%%%%%%%%%%%%%%%%%%%%%%%%%%%%%%%%%%%%%%%%
	\begin{align}
	\Psi^{(C)}_{R/L,\pm}\sim \Psi_{L/R,\mp}^{{\ast}}.
	\end{align}
	%%%%%%%%%%%%%%%%%%%%%%%%%%%%%%%%%%%%%%%%%%%%%%%%%%%%%%%%%%%%%%
Since {components with the same 4d chiralities (but opposite inner chiralities) are related by} the ${\cal C}$ transformation, the 6d charge conjugation $C$ itself cannot be {the origin of the} ${\cal C}$ transformation. 
Here, we focus on \cred{the fact} that the parity transformation $P$,\,\footnote{The gamma matrix $\Gamma^{y}$ in the parity transformation (\ref{6dparity}) plays \cred{the} role of the $\pi$-rotation in the $y_{1}y_{2}$-plane. The ${\cal C}$ transformation does not change the sign of the extra dimension coordinates\cred{;} we multiplied $\Gamma^{y}$ instead of \cred{the} replacement $y \to -y$.}
	%%%%%%%%%%%%%%%%%%%%%%%%%%%%%%%%%%%%%%%%%%%%%%%%%%%%%%%%%%%%%%
	\begin{align}
	P:\Psi(t,{\bm x},y)\rightarrow\ \Psi^{(P)}(t,{\bm x},y)=\Gamma^{0}\Gamma^{y}\Psi(t,-{\bm x},y), \label{6dparity}
	\end{align}
	%%%%%%%%%%%%%%%%%%%%%%%%%%%%%%%%%%%%%%%%%%%%%%%%%%%%%%%%%%%%%%
flips only the 4d chirality $R/L$ as 
	%%%%%%%%%%%%%%%%%%%%%%%%%%%%%%%%%%%%%%%%%%%%%%%%%%%%%%%%%%%%%%
	\begin{align}
	\Psi^{(P)}_{R/L,\pm}\sim \Psi_{L/R,\pm},
	\end{align}
	%%%%%%%%%%%%%%%%%%%%%%%%%%%%%%%%%%%%%%%%%%%%%%%%%%%%%%%%%%%%%%
so that the 6d CP transformation, which is {the} combination of the 6d charge conjugation $C$ and the parity transformation $P$, flips {only} the inner chirality $\pm$ and can {correspond to} the ${\cal C}$ transformation,\,\footnote{
{Note that this CP transformation is not equal to the ``modified'' CP transformation which is useful for discussing CP violation \cred{from} the 4d point of view~\cite{Lim:1990bp,Lim:2009pj,Kobayashi:2016qag} in $4+2n\,(n=1,2,\cred{\ldots})$ dimensions.} 
}
	%%%%%%%%%%%%%%%%%%%%%%%%%%%%%%%%%%%%%%%%%%%%%%%%%%%%%%%%%%%%%%
	\begin{align}
	{CP}:\Psi(t,{\bm x},y)\rightarrow\ \Psi^{(CP)}(t,{\bm x},y)=\Gamma^{0}\Gamma^{y} {C} \overline{\Psi}^{T}(t,-{\bm x},y).\label{6dCP}
	\end{align}
	%%%%%%%%%%%%%%%%%%%%%%%%%%%%%%%%%%%%%%%%%%%%%%%%%%%%%%%%%%%%%%
In fact, \cred{multiplying} $\Gamma^{5}$ and $\Gamma^{y}$ and using the properties of the 6d charge conjugation given in \cred{Appendix}~\ref{sec:appendixB}, we can easily check that the 6d CP transformation only flips the inner chirality $\pm$:
	%%%%%%%%%%%%%%%%%%%%%%%%%%%%%%%%%%%%%%%%%%%%%%%%%%%%%%%%%%%%%%
	\begin{align}
	\Psi^{(CP)}_{R/L,\pm}\sim \Psi_{R/L,\mp}^{{\ast}}{.}
	\end{align}
	%%%%%%%%%%%%%%%%%%%%%%%%%%%%%%%%%%%%%%%%%%%%%%%%%%%%%%%%%%%%%%
We should mention that the action (\ref{Dirac-action}) is invariant under the 6d CP transformation (\ref{6dCP}) and the CP-transformed Dirac fermion $\Psi^{(CP)}(t,{\bm x},y)$ satisfies the same 6d Dirac equation (\ref{Dirac-equation}) as the original Dirac fermion $\Psi (x,y)$.
This implies that the 6d CP transformation does not change the spectrum and could connect the \cred{degenerate} solutions of the Dirac equation if {they} exist, as the ${\cal C}$ transformation.
In the \cred{chiral} representation {of 6d Gamma matrices} (\cred{see Appendix~\ref{sec:appendixA} for detail}), the 6d CP transformation {(\ref{6dCP})} {is represented in} the following concrete form \cred{by} regarding $\xi_{R/L,\pm}(x,y)$ as two-{component} spinors:
 	%%%%%%%%%%%%%%%%%%%%%%%%%%%%%%%%%%%%%%%%%%%%%%%%%%%%%%%%%%%%%%
	\begin{align}
	\left(\begin{array}{c}
		\xi^{(CP)}_{R+}\\[0.2cm]
		\xi^{(CP)}_{L+}\\[0.2cm]
		\xi^{(CP)}_{R-}\\[0.2cm]
		\xi^{(CP)}_{L-}
		\end{array}\right)(t,{\bm x},y)=C\left(\begin{array}{c}
		\xi_{R+}\\[0.2cm]
		\xi_{L+}\\[0.2cm]
		-\xi_{R-}\\[0.2cm]
		-\xi_{L-}
		\end{array}\right)^{\ast}(t,-{\bm x},y),\label{6dCPweyl}
	\end{align}
	%%%%%%%%%%%%%%%%%%%%%%%%%%%%%%%%%%%%%%%%%%%%%%%%%%%%%%%%%%%%%%
where
	%%%%%%%%%%%%%%%%%%%%%%%%%%%%%%%%%%%%%%%%%%%%%%%%%%%%%%%%%%%%%%
	\begin{align}
	C&=i\sigma_{2}\otimes C^{(\cred{\rm 4d})}.
	\end{align}
	%%%%%%%%%%%%%%%%%%%%%%%%%%%%%%%%%%%%%%%%%%%%%%%%%%%%%%%%%%%%%%
$C^{(\cred{\rm 4d})}=i\gamma^{2}\gamma^{0}$ is the ({ordinary}) 4d charge conjugation.
{We can see from (\ref{6dCPweyl}) that} the 6d CP transformation contains the 4d CP transformation to connect $\Psi_{R,+}$ ($\Psi_{L,+}$) and $\Psi_{R,-}$ ($\Psi_{L,-}$) without changing the 4d chirality as the 4d CP transformation.
{In the basis defined in Eq.~(\ref{eq:new_basis_fermion}),} {rearranging the order of the components in Eq.~(\ref{6dCPweyl}), we can rewrite it \cred{in} the form}
 	%%%%%%%%%%%%%%%%%%%%%%%%%%%%%%%%%%%%%%%%%%%%%%%%%%%%%%%%%%%%%%
	\begin{align}
	\left(\begin{array}{c}
		{-}\xi^{(CP)}_{R+}\\[0.2cm]
		\xi^{(CP)}_{R-}\\[0.2cm]
		\xi^{(CP)}_{L+}\\[0.2cm]
		\xi^{(CP)}_{L-}
		\end{array}\right)(t,{\bm x},y)=
								\left[
								\left(\begin{array}{cc}
								\sigma_{1}&0\\
								0&-\sigma_{1}
								\end{array}\right)
								{\otimes (-{{\rm I}_2})}
								\right]
								\left(\begin{array}{c}
		{i}\sigma_{2}\xi_{R+}^{\ast}\\[0.2cm]
		{i}\sigma_{2}(-\xi_{R-})^{\ast}\\[0.2cm]
		{-i}\sigma_{2}(\xi_{L+})^{\ast}\\[0.2cm]
		{-i}\sigma_{2}(\xi_{L-})^{\ast}
		\end{array}\right)(t,-{\bm x},y),
	%\label{6dCPweyl}
	\end{align}
	%%%%%%%%%%%%%%%%%%%%%%%%%%%%%%%%%%%%%%%%%%%%%%%%%%%%%%%%%%%%%%
{where $i\sigma_{2}\xi_{R\pm}^{\ast}$ and $-i\sigma_{2}\xi_{L\pm}^{\ast}$ are CP-transformed fields in the 4d sense.}
{The above transformation with respect to the extra dimensions is found to correspond to the {$\mathcal{C}$} transformation (\ref{Ctransformation}).}
Thus, we can understand that the {$4 \times 4$ ${\cal C}$ matrix originates from} the 6d CP transformation,
{where $(-{{\rm I}_2})$ shows the trivial rotation of two-component spinors with an unphysical overall minus sign.}

%%%%%%%%%%%%%%%%%%%%%%%%%%%%%%%%%%%%%
%%%%%%%% Section 8 %%%%%%%%%%%%%%%%%%
%%%%%%%%%%%%%%%%%%%%%%%%%%%%%%%%%%%%%
\section{\cred{Six-dimensional} Dirac fermion on arbitrary flat surfaces with boundaries
\label{sec:general_BC}}

\begin{figure}[t]
\centering
\includegraphics[clip, width=0.50\hsize]{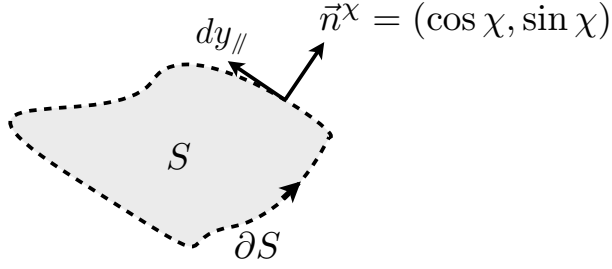}
\caption{$\partial S$ denotes the boundary of $S$.
$\vec{n}^\chi$ is a unit normal vector orthogonal to $\partial S$, and $d y_{\varParallel}$ is a line element along the boundary $\partial S$.}
\label{fig:boundary}
\end{figure}

So far, we have restricted our considerations to the rectangle as a two-dimensional extra space.
For phenomenological applications, it will be useful to extend our analysis to arbitrary flat surfaces $S$ with boundaries, like polygons, a disk, etc.
To this end, we introduce the inner product for four-component wavefunctions $\Phi'(y)$ and $\Phi(y)$ on $S$ as 
\begin{align}
\langle \Phi', \Phi \rangle_S = \int_S dy_1 dy_2 \left( \Phi'(y) \right)^\dagger \Phi(y).
\end{align}
The requirement \cred{is} that the supercharge $Q$ is given by
\begin{align}
\langle Q \Phi', \Phi \rangle_S = \langle \Phi', Q \Phi \rangle_S.
	\label{eq:Q_condition_general}
\end{align}
By expressing the supercharge $Q$ {defined in Eq.~(\ref{supercharge})} \cred{in} the form
\begin{align}
Q = i \partial_{y_j} \widetilde{\Gamma}_j + M \widetilde{\Gamma}_M\qquad
(j=1,2)
\end{align}
with
\begin{align}
\widetilde{\Gamma}_1 = \begin{pmatrix} 0 & -\sigma_2 \\ -\sigma_2 & 0 \end{pmatrix},\quad
\widetilde{\Gamma}_2 = \begin{pmatrix} 0 &  \sigma_1 \\  \sigma_1 & 0 \end{pmatrix},\quad
\widetilde{\Gamma}_M = \begin{pmatrix} 0 &  \sigma_3 \\  \sigma_3 & 0 \end{pmatrix},
\end{align}
we have found that the condition~(\ref{eq:Q_condition_general}) leads to
\begin{align}
\oint_{\partial S} dy_{\varParallel} \left( \Phi'(y) \right)^\dagger \left( n^\chi_j \widetilde{\Gamma}_j \right) \Phi(y) = 0,
	\label{eq:non-local_equation}
\end{align}
where $\partial S$ denotes the boundary of the surface $S$, $(n^\chi_1, n^\chi_2) = (\cos{\chi}, \sin{\chi})$ is a unit normal vector orthogonal to the boundary $\partial S$, and $d y_{\varParallel}$ is a line element along $\partial S$, as depicted in Fig.~\ref{fig:boundary}.

Since it is hard to solve the non-local equation~(\ref{eq:non-local_equation}) in general, we will here restrict our considerations to the case that the \cred{local condition}
\begin{align}
\left( \Phi'(y) \right)^\dagger \left( n^\chi_j \widetilde{\Gamma}_j \right) \Phi(y) = 0\qquad
\text{at } (y_1, y_2) \in \partial S
	\label{eq:local_equation}
\end{align}
is satisfied at each point of the boundary $\partial S$, as was done in section~\ref{sec:Classification of allowed boundary conditions}.

Although the condition~(\ref{eq:local_equation}) should be satisfied for arbitrary four-component wavefunctions $\Phi'(y)$ and $\Phi(y)$, it is actually sufficient to solve Eq.~(\ref{eq:local_equation}) for $\Phi'(y) = \Phi(y)$, as was shown in section~\ref{sec:Classification of allowed boundary conditions}.
Inserting
\begin{align}
\left( n_1^\chi, n_2^\chi \right) = \left( \cos{\chi}, \sin{\chi} \right)
\end{align}
into~(\ref{eq:local_equation}) with $\Phi'(y) = \Phi(y) = (f_1(y), f_2(y), g_1(y), g_2(y)^{\rm T}$ leads to
\begin{align}
0 &= \left( \Phi'(y) \right)^\dagger \left( n^\chi_j \widetilde{\Gamma}_j \right) \Phi(y) \notag \\
  &= \left( \rho(y) \right)^\dagger \sigma_\chi \lambda(y) + \left( \sigma_\chi \lambda(y) \right)^\dagger \rho(y),
  \label{eq:condition_general}
\end{align} 
where
\begin{align}
\rho(y) \equiv \begin{pmatrix} f_1(y) \\ f_2(y) \end{pmatrix},\quad
\lambda(y) \equiv \begin{pmatrix} g_1(y) \\ g_2(y) \end{pmatrix}
\end{align}
and
\begin{align}
\sigma_\chi \equiv - \cos{\chi} \, \sigma_2 + \sin{\chi} \, \sigma_1 = \left( \sigma_\chi \right)^\dagger.
\end{align}
A crucial observation is that the condition~(\ref{eq:condition_general}) can be rewritten as
\begin{align}
\left |\rho(y) + L_0 \sigma_\chi \lambda(y) \right|^2 =
\left |\rho(y) - L_0 \sigma_\chi \lambda(y) \right|^2,
	\label{eq:condition_general_2}
\end{align}
where $L_0$ is a non-zero real constant whose value is irrelevant unless $L_0$ is non-vanishing.

General solutions to~(\ref{eq:condition_general_2}) are easily found in the form
\begin{align}
\rho(y) + L_0 \sigma_\chi \lambda(y) = U
\left( \rho(y) - L_0 \sigma_\chi \lambda(y) \right),
\end{align}
or equivalently
\begin{align}
({\rm I}_2 - U) \rho(y) = - L_0 ({\rm I}_2 + U) \sigma_\chi \lambda(y),
	\label{eq:condition_general_3}
\end{align}
where $U$ is an arbitrary two-by-two unitary matrix.
Following the arguments given in section~\ref{sec:Classification of allowed boundary conditions}, we conclude that the condition~(\ref{eq:condition_general_3}) has to reduce to
\begin{align}
({\rm I}_2 - U) \rho(y) &= 0,               \label{eq:BCorigin_general_1} \\
({\rm I}_2 + U) \sigma_\chi \lambda(y) &= 0,   \label{eq:BCorigin_general_2}
\end{align}
and\cred{, further,} that the allowed boundary conditions are classified into three types:
%%%%%%%
\begin{description}
%%%%%%%%%%%%%%%%%%%%%%%%%%%%%%%%%%%%%%%%%%%%%%%%%%%%%%%%%%%%%%%
\item[\r{1})] \ \underline{Type I boundary condition}\\[0.3cm]
%%%%%%%%%%%%%%%%%%%%%%%%%%%%%%%%%%%%%%%%%%%%%%%%%%%%%%%%%%%%%%%
\begin{align}
U_{\rm Type \ I} = \begin{pmatrix} 1 & 0 \\ 0 & 1 \end{pmatrix}.
\end{align}
It follows that the condition~(\ref{eq:BCorigin_general_1}) is trivially satisfied, and the condition~(\ref{eq:BCorigin_general_2}) reduces to
\begin{align}
g_1(y) = g_2(y) = 0 \qquad \text{at } (y_1, y_2) \in \partial S.
	\label{eq:generalBC_I_1}
\end{align}
It will be convenient to rewrite the boundary condition~(\ref{eq:generalBC_I_1}), in terms of the original four-component wavefunction $\Phi(y)$ \cred{as}
\begin{align}
{\cal P}_{(-1)^F = -1} \Phi(y) = 0 \qquad \text{at } (y_1, y_2) \in \partial S,
\end{align}
with
\begin{align}
{\cal P}_{{(-1)^F = \pm1}} = \frac{1}{2} \left( {\rm I}_4 \pm (-1)^F \right).
\end{align}
%%%
%%%%%%%%%%%%%%%%%%%%%%%%%%%%%%%%%%%%%%%%%%%%%%%%%%%%%%%%%%%%%%%
\item[\r{2})] \ \underline{Type \GkII \ boundary condition}\\[0.3cm]
%%%%%%%%%%%%%%%%%%%%%%%%%%%%%%%%%%%%%%%%%%%%%%%%%%%%%%%%%%%%%%%
\begin{align}
U_{\rm Type \ \GkII} = \begin{pmatrix} -1 & 0 \\ 0 & -1 \end{pmatrix}.
\end{align}
It follows that the condition~(\ref{eq:BCorigin_general_2}) is trivially satisfied, while the condition~(\ref{eq:BCorigin_general_1}) reduces to
\begin{align}
f_1(y) = f_2(y) = 0 \qquad \text{at } (y_1, y_2) \in \partial S.
\end{align}
In \cred{terms} of $\Phi(y)$, the above boundary condition can be expressed as
\begin{align}
{\cal P}_{(-1)^F = +1} \Phi(y) = 0 \qquad \text{at } (y_1, y_2) \in \partial S.
\end{align}
%%%
%%%%%%%%%%%%%%%%%%%%%%%%%%%%%%%%%%%%%%%%%%%%%%%%%%%%%%%%%%%%%%%
\item[\r{3})] \ \underline{Type \GkIII \ boundary condition}\\[0.3cm]
%%%%%%%%%%%%%%%%%%%%%%%%%%%%%%%%%%%%%%%%%%%%%%%%%%%%%%%%%%%%%%%
\begin{align}
U_{\rm Type \ \GkIII} = \vec{n} \cdot \vec{\sigma} =
\begin{pmatrix} \cos\theta & e^{-i\phi} \sin\theta \\ e^{i\phi} \sin\theta & -\cos\theta \end{pmatrix},
	\label{eq:U_type-III}
\end{align}
with
\begin{align}
\vec{n} = \left( \cos\phi \sin\theta, \sin\phi \sin\theta, \cos\theta \right).
\end{align}
It follows from~(\ref{eq:U_type-III}) that Eqs.~(\ref{eq:BCorigin_general_1}) and (\ref{eq:BCorigin_general_2}) become
\begin{align}
({\rm I}_2 - \vec{n} \cdot \vec{\sigma}) \rho(y) =
({\rm I}_2 - \vec{n} \cdot \vec{\sigma}) \begin{pmatrix} f_1(y) \\ f_2(y) \end{pmatrix} &= 0, \notag \\
({\rm I}_2 + \vec{n} \cdot \vec{\sigma'})\lambda(y) =
({\rm I}_2 + \vec{n} \cdot \vec{\sigma'}) \begin{pmatrix} g_1(y) \\ g_2(y) \end{pmatrix} &= 0\qquad \text{at } (y_1, y_2) \in \partial S,
\end{align}
with
\begin{align}
\vec{\sigma'} &\equiv \sigma_\chi \vec{\sigma} \sigma_\chi \notag \\
&= (-\cos(2\chi)\sigma_1-\sin(2\chi)\sigma_2, \cos(2\chi)\sigma_2-\sin(2\chi)\sigma_1, -\sigma_3).
\end{align}
{Here, we used the property $(\sigma_\chi)^2 = {\rm I}_2$.}
It will be convenient to express the above boundary condition in terms of the original four-component wavefunction $\Phi(y)$.
The result is given by
\begin{align}
{\cal P}_{\vec{n} \cdot \vec{\Sigma} = -1} \Phi(y) = 0 \qquad \text{at } (y_1, y_2) \in \partial S,
\end{align}
where ${\cal P}_{\vec{n} \cdot \vec{\Sigma} = \pm 1}$ are projection matrices defined by
\begin{align}
{\cal P}_{\vec{n} \cdot \vec{\Sigma} = \pm 1} &\equiv 
\frac{1}{2} \left( {\rm I}_4 \pm \vec{n} \cdot \vec{\Sigma} \right), \\
\vec{\Sigma} &\equiv \begin{pmatrix} \vec{\sigma} & 0 \\ 0 & -\vec{\sigma'} \end{pmatrix}.
\end{align}
%%
%%%%%%%%%%%%%%%%
\end{description}

We have succeeded \cred{in classifying} the allowed boundary conditions at each point of the boundary $\partial S$.
We should note that the results given in this section are consistent with those in section~\ref{sec:Classification of allowed boundary conditions}.
Actually, for $\chi = \pm \pi$ ($\chi = \pm \pi/2$), the above results {reduce} to those given in the subsection~\ref{sec:Allowed boundary conditions in the y_{1}-direction} (\ref{sec:Allowed boundary conditions in the y_{2}-direction}).

Let us examine an $n$-sided polygon as an application of \cred{the} analysis given above.
Let $\vec{n}^{\chi_a} = (\cos\chi_a, \sin\chi_a)\ (a=1,2,\cred{\ldots},n)$ be a normal unit vector orthogonal to the \cred{$a$th} side of the polygon.
Then, we can impose one of the following boundary conditions on the \cred{$a$th} side of the polygon: 
\begin{align}
  {\rm Type \ I}:&\  {\cal P}_{(-1)^F = -1} \Phi(y) = 0, \notag \\
 {\rm Type \ \GkII}:&\  {\cal P}_{(-1)^F = +1} \Phi(y) = 0, \notag \\
{\rm Type \ \GkIII}:&\ \ \, {\cal P}_{\vec{n} \cdot \vec{\Sigma}_a = -1} \Phi(y) = 0,
\end{align}
with
\begin{align}
\vec{\Sigma}_a &= \begin{pmatrix} \vec{\sigma} & 0 \\ 0 & - \sigma_{\chi_a} \vec{\sigma} \sigma_{\chi_a} \end{pmatrix}{.}
\end{align}
%on the $a$-th side of the polygon.
{If} we would like to impose a single boundary condition on every side of the polygon, the possible boundary conditions are restricted to
\begin{align}
\cred{(1)}&\ g_1(y) = g_2(y) = 0, \notag \\
\cred{(2)}&\ f_1(y) = f_2(y) = 0, \notag \\
\cred{(3)}&\ f_1(y) = g_1(y) = 0, \notag \\
\cred{(4)}&\ f_2(y) = g_2(y) = 0,
	\label{eq:four_BCs}
\end{align}
on every side of the polygon.
The above boundary conditions \cred{$(1)$, $(2)$, $(3)$ and $(4)$ correspond to Type I, Type $\text{\GkII}$, Type $\text{\GkIII}$ with $\theta = \pi$, and Type $\text{\GkIII}$ with $\theta = 0$}, respectively.
We note that the allowed Type $\text{\GkIII}$ boundary conditions are limited to {$\theta = \pi$ and $0$, where $\phi$ does not contribute to the boundary conditions at $\theta = \pi$ and $0$.}
This is because the normal unit vector $\vec{n}^{\chi_a} \ (a=1,2,\cred{\ldots},n)$ on the \cred{$a$th} side is independent of $\vec{n}^{\chi_b}$ for $a \not= b$, in general, so that ${\cal P}_{\vec{n} \cdot \vec{\Sigma}_a = -1} \ (a=1,2,\cred{\ldots},n)$ cannot be identical for all sides of the polygon expect for {$\theta = \pi$ and $0$, irrespective of $\phi$}.

Let us finally discuss a disk as the extra dimensions.
For a disk, we may impose a single boundary condition on every point of the edge of the disk.
It then follows from the analysis of the $n$-sided polygon that the boundary condition on the edge of the disk has to be chosen from one of \cred{the} four boundary conditions~(\ref{eq:four_BCs}), otherwise the \cred{Hermiticity} of the supercharge would be lost.

%%%%%%%%%%%%%%%%%%%%%%%%%%%%%%%%%%%%%
%%%%%%%% Section 9 %%%%%%%%%%%%%%%%%%
%%%%%%%%%%%%%%%%%%%%%%%%%%%%%%%%%%%%%
\section{Conclusions and discussions
\label{sec:Conclusions and discussions}}

We have succeeded in revealing the supersymmetric structure hidden in the 6d Dirac action on a 
rectangle. The supersymmetry turns out to be very useful to classify the class of allowed 
boundary conditions, and to clarify the 4d mass spectrum of the \cred{Kaluza--Klein} modes for the 6d 
Dirac fermion. In fact, the allowed boundary conditions are derived by demanding the \cred{Hermiticity}
of the supercharge and are classified into three types.
{We have furthermore extended our analysis to arbitrary flat surfaces
as the two-dimensional extra space. We have then found that the
supersymmetric structure is still realized there and \cred{have} succeeded in
classifying the allowed boundary conditions, in general.}

An important observation in our results is that two massless chiral fermions appear in the 
4d mass spectrum for \cred{the} {Type \GkI} or {Type \GkII} boundary \cred{conditions}. This result seems to be
surprising because the 6d Dirac fermion is non-chiral and furthermore has the non-vanishing 
bulk mass $M$.\footnote{
{It should be emphasized that no \cred{zero-energy} solution or no 4d massless chiral fermion \cred{appears} for the non-vanishing bulk mass $M$ if we take the torus as the two-dimensional extra space, instead of the rectangle.}}
Then, one might naively expect that the 4d mass spectrum would consist of 
only massive states with masses heavier than $M$. Actually, \cred{positive-energy} eigenstates
correspond to massive 4d Dirac fermions with masses 
$m_{n_{1}n_{2}}>M$ for $n_{1}, n_{2}=1,2,3,\cred{\ldots}$.

On the other hand, we have found that the 4d massless chiral fermions correspond to \cred{zero-energy} solutions, which are bound states and possess a topological nature in supersymmetric 
quantum mechanics. The appearance of the degenerate 4d massless chiral fermions will become 
crucially important \cred{in solving} the generation problem and also the fermion mass hierarchy \cred{problem} 
of the quarks and leptons, though the 4d massless chiral fermions are two-fold degenerate 
but not \cred{three-} in the present 6d model.

In our analysis, we have found \cred{the} remarkable feature that \cred{zero-energy} solutions are not 
affected by the presence of the boundaries, while the boundary conditions work well for 
determining the \cred{positive-energy} solutions. Even though we have explicitly constructed a 
one-parameter family of the \cred{zero-energy} solutions (5.18) and (5.22) for the \cred{Type} {\GkII} \cred{boundary condition} 
and shown that the number of the degeneracy is two, the analysis seems to be insufficient. 
This is because \cred{the} general class of the \cred{zero-energy} solutions is much wider than
\cred{considered here}, and we have not succeeded in determining a complete set of \cred{zero-energy solutions definitely}.\footnote{{
It is worth \cred{noting} that no trouble appears in 5d fermion systems with a single extra 
dimension, though a similar situation happens there \cite{Fujimoto:2011kf, Fujimoto:2012wv, 
Fujimoto:2013ki, Fujimoto:2014fka}. Any \cred{zero-energy} solution is not degenerate in one dimension, 
so that it can be determined uniquely.}}
Since \cred{zero-energy} solutions are directly related to massless 4d chiral fermions, it would 
be of great importance to clarify the structure of the \cred{zero-energy} solutions for 
higher-dimensional Dirac systems with more than or equal to  \cred{two extra} dimensions, 
phenomenologically as well as mathematically.\footnote{
\cred{Determining} the size and the shape of \cred{the} extra dimensions, known as moduli stabilization, would be \cred{issues closely related} to gravitational effects in higher-dimensional \cred{space-time}, which is absent in the present flat setup.
Though this subject is of importance for a complete discussion on models in the context of extra dimension, we will
leave it for a topics for future studies\cred{.}
{Another extension is to consider curved extra dimensions. Even for this situation, the supersymmetric structure is expected to be realized~\cite{Lim:2007fy,Nagasawa:2011mu}. It would be of interest to study the above subjects.}}

\cred{One extension of our analysis} is to introduce 
potential terms in the Hamiltonian. This can be done by replacing the bulk mass $M$ by a 
superpotential $W(y)$ in the supercharge {$Q$} {in} (\ref{supercharge}). Even with the
superpotential $W(y)$, the supercharge is still \cred{Hermitian} for {Type} I, {\GkII} and 
{\GkIII} boundary conditions. Interestingly, the superpotential may naturally be introduced 
through a Yukawa interaction $g\langle\phi (y)\rangle\overline{\Psi}(x,y)\Psi (x,y)$ with 
a non-trivial background $\langle\phi (y)\rangle$ of {a} scalar field $\phi(x,y)$.

Another important extension of our analysis is to investigate higher-dimensional Dirac
actions.
In the case of {a 6d Dirac fermion}, only two massless chiral fermions appear in the 4d mass spectrum, which is not sufficient to solve the generation problem.
However, more than two 4d massless chiral fermions may 
appear in the case of higher dimensions, equal to \cred{or more than eight} dimensions,
%
%
%%%%%%%%% New! added by sakamoto %%%%%%%%%%%%
even though it is \cred{naively expected} that $2^{n}$ massless chiral fermions
would appear in the case of $D=4+2n$.
This may imply that it is very important to perform a comprehensive analysis
of \cred{the} allowed boundary conditions in higher-dimensional Dirac actions, 
as done in this paper, because a suitable choice of boundary conditions
could reduce \cred{the possible} $2^{n}$ massless chiral fermions to \cred{three} massless ones.
Thus, {it} would be of great interest to extend our analysis to 
higher-dimensional Dirac fermions and \cred{to search for the possibility of producing}
a three-generation model.
\cred{This} work will be reported elsewhere.
%%%%%%%%%%%%%%%%%%%%%%%%%%%%%%%%%%%%%%%%%%%%%
%
%%%%%%%%% remo\ved by sakamoto %%%%%%%%%%%
%. It would be expected naively that $2^{n}$ massless chiral fermions appear in the case of $D=4+2n$ , however, a suitable choice of boundary conditions from a general class, which would be obtained by using the same technique of this paper, may give us a possibility to get a three-generation model. Arbitrary higher-dimensional extension of the analysis will be published elsewhere and the interesting models on 
%extra dimension in phenomenological sense},
%which naturally solve the problems of fermion generation and mass hierarchy, will be reported elsewhere.
%In particular, it would be of great importance to analyze Dirac actions in 
%%%%%%%%%%%%%%%%%%%%%%%%%%%%%%%%%%%%%%%%%%

\section*{\cred{Acknowledgments}}

We thank Tomoaki Nagasawa for discussions in the early \cred{stages} of this work.
This work is supported in part by Grants-in-Aid for Scientific 
Research [No.~15K05055 and No.~25400260 (M.S.)] from the Ministry of Education, 
Culture, Sports, Science and Technology (MEXT) in Japan.

\appendix
\section*{Appendix}
\section{Chiral representation {of 6d Gamma matrices}}\label{sec:appendixA}
In this appendix, we represent our choice of the \cred{chiral} representation {of the 6d Gamma matrices}:
	%%%%%%%%%%%%%%%%%%%%%%%%%%%%%%%%%%%%%%%%
	\begin{align}
	\Gamma^{\mu}&={{\rm I}_2}\,\otimes\gamma^{\mu}=\left(\begin{array}{cc}
													\gamma^{\mu}&0\\
													0&\gamma^{\mu}
													\end{array}\right)=\left(\begin{array}{c:c}
						\begin{array}{cc}
						0&\sigma^{\mu}\\
						\overline{\sigma}^{\mu}&0						
						\end{array}& \\
						\hdashline
						&\begin{array}{cc}
						0&\sigma^{\mu}\\
						\overline{\sigma}^{\mu}&0
						\end{array}
					\end{array}\right),\label{gammamuw}\\[0.2cm]
	\Gamma^{y_{1}}&=i\sigma_{1}\otimes \gamma^{5}=\left(\begin{array}{cc}
													0&i\gamma^{5}\\
													i\gamma^{5}&0
													\end{array}\right)=\left(\begin{array}{c:c}
						&\begin{array}{cc}
						i{{\rm I}_2}& 0\\
						0&-{i{\rm I}_2}						
						\end{array}\\
						\hdashline
						\begin{array}{cc}
						i{{\rm I}_2}&0\\
						0&-i{{\rm I}_2}
						\end{array}&
					\end{array}\right),\label{gammay1w}\\[0.2cm]
	\Gamma^{y_{2}}&=i\sigma_{2}\otimes \gamma^{5}=\left(\begin{array}{cc}
													0&\gamma^{5}\\
													-\gamma^{5}&0
													\end{array}\right)=\left(\begin{array}{c:c}
						&\begin{array}{cc}
						{{\rm I}_2}& 0\\
						0&-{{\rm I}_2}						
						\end{array}\\
						\hdashline
						\begin{array}{cc}
						-{{\rm I}_2}&0\\
						0&{{\rm I}_2}
						\end{array}&
					\end{array}\right),\label{gammay2w}
	\end{align}
	%%%%%%%%%%%%%%%%%%%%%%%%%%%%%%%%%%%%%%%%
{with $\sigma^\mu = ({\bm 1}_2, -\sigma_1, -\sigma_2, -\sigma_3)$ and $\bar{\sigma}^\mu = ({\bm 1}_2, \sigma_1, \sigma_2, \sigma_3)$.}
In this basis, the 4d chirality and the inner chirality are expressed with the following diagonal forms:
	%%%%%%%%%%%%%%%%%%%%%%%%%%%%%%%%%%%%%%%%	
	\begin{align}
	\Gamma^{5}&\equiv i\Gamma^{0}\Gamma^{1}\Gamma^{2}\Gamma^{3}={{\rm I}_2}\otimes (i\gamma^{0}\gamma^{1}\gamma^{2}\gamma^{3})\nonumber\\
	&={{\rm I}_2}\otimes \gamma^{5}=\left(\begin{array}{cc}
													\gamma^{5}&0\\
													0&\gamma^{5}
													\end{array}\right)=\left(\begin{array}{c:c}
						\begin{array}{cc}
						{{\rm I}_2}& 0\\
						0&-{{\rm I}_2}						
						\end{array}&\\
						\hdashline
						&\begin{array}{cc}
						{{\rm I}_2}&0\\
						0&-{{\rm I}_2}
						\end{array}
					\end{array}\right),\label{gamma5w}\\[0.2cm]
	\Gamma^{y}&\equiv i \Gamma^{y_{1}}\Gamma^{y_{2}}\nonumber\\
		&=\sigma_{3}\otimes {{\rm I}_4}=\left(\begin{array}{cc}
													{{\rm I}_4}&0\\
													0&-{{\rm I}_4}
													\end{array}\right)=\left(\begin{array}{c:c}
						\begin{array}{cc}
						{{\rm I}_2}& 0\\
						0&{{\rm I}_2}						
						\end{array}&\\
						\hdashline
						&\begin{array}{cc}
						-{{\rm I}_2}&0\\
						0&-{{\rm I}_2}
						\end{array}
					\end{array}\right),\label{gammay3w}
	\end{align}
	%%%%%%%%%%%%%%%%%%%%%%%%%%%%%%%%%%%%%%%%
{As a result,} \cred{the} {eight-component spinors} $\Psi_{R/L,\pm}$, which {are} simultaneous eigenstates of $\Gamma^{5}$ and $\Gamma^{y}$, {are} expressed {in terms of} {\cred{two-component} spinors} $\xi_{R/L,\pm}$ as
	%%%%%%%%%%%%%%%%%%%%%%%%%%%%%%%%%%%%%%%%
	\begin{align}
	\Psi_{R,+}=\left(\begin{array}{c}
				\xi_{R,+}\\
				0\\
				0\\
				0
				\end{array}\right),\quad
	\Psi_{L,+}=\left(\begin{array}{c}
				0\\
				\xi_{L,+}\\
				0\\
				0
				\end{array}\right),\quad
	\Psi_{R,-}=\left(\begin{array}{c}
				0\\
				0\\
				\xi_{R,-}\\
				0
				\end{array}\right),\quad
	\Psi_{L,-}=\left(\begin{array}{c}
				0\\
				0\\
				0\\
				\xi_{L,-}
				\end{array}\right).	
	\end{align}
	%%%%%%%%%%%%%%%%%%%%%%%%%%%%%%%%%%%%%%%%
\section{\cred{Six-dimensional} charge conjugation}\label{sec:appendixB}
In this appendix, we \cred{show} the definition of the 6d charge conjugation\cred{, $C$:}
	%%%%%%%%%%%%%%%%%%%%%%%%%%%%%%%%%%%%%%%%
	\begin{align}
	C:\Psi(x,y)\rightarrow\ \Psi^{(C)}(x,y)&=C\overline{\Psi}{}^{{\rm T}}(x,y)\nonumber\\
	&=C(\Gamma^{0})^{{\rm T}}\Psi^{\ast}(x,y),\label{6dC}
	\end{align}
	%%%%%%%%%%%%%%%%%%%%%%%%%%%%%%%%%%%%%%%%
	In \cred{the} 6d case, the charge conjugation matrix $C$ satisfies the following relations:
	%%%%%%%%%%%%%%%%%%%%%%%%%%%%%%%%%%%%%%%%
	\begin{align}
	&C^{-1}\Gamma^{M}C= - (\Gamma^{M})^{{\rm T}},\\
	&C^{\dagger}C={{\rm I}_8},\\
	&C^{{\rm T}}=C.
	\end{align}
	%%%%%%%%%%%%%%%%%%%%%%%%%%%%%%%%%%%%%%%%	
In general, we have two choices for 6d charge conjugation:
	%%%%%%%%%%%%%%%%%%%%%%%%%%%%%%%%%%%%%%%%
	\begin{align}
	&C_{\eta}^{-1}\Gamma^{M}C_{\eta}= \eta (\Gamma^{M})^{{\rm T}},\\
	&C_{\eta}^{\dagger}C_{\eta}={{\rm I}_8},\\
	&C^{{\rm T}}_{\eta}=-\eta^{3}C_{\eta},\qquad (\eta=\pm1).
	\end{align}
	%%%%%%%%%%%%%%%%%%%%%%%%%%%%%%%%%%%%%%
%
%
{For concrete discussions, we adopt the following form}
%%%%%%%%% New! added by sakamoto %%%%%%%%%%%%
%
	%%%%%%%%%%%%%%%%%%%%%%%%%%%%%%%%%%%%%%%%
	\begin{align}
	C = i\sigma_{2} \otimes C^{(\cred{\rm 4d})},
	\notag
	\end{align}
	%%%%%%%%%%%%%%%%%%%%%%%%%%%%%%%%%%%%%%%%
%
where ${C^{(\cred{\rm 4d})}} = i\gamma^{2}\gamma^{0}$ is the 4d charge conjugation matrix.
%
%%%%%%%%% removed by sakamoto %%%%%%%%%%%%
%\textcolor{red}{		
%In general, we have two choices for 6d charge conjugation:
%	%%%%%%%%%%%%%%%%%%%%%%%%%%%%%%%%%%%%%%%%
%	\begin{align}
%	&C_{\eta}^{-1}\Gamma^{M}C_{\eta}= \eta (\Gamma^{M})^{T},\\
%	&C_{\eta}^{\dagger}C_{\eta}={\bm 1}_{8},\\
%	&C^{t}_{\eta}=-\eta^{3}C_{\eta},\qquad (\eta=\pm1).
%	\end{align}
%	%%%%%%%%%%%%%%%%%%%%%%%%%%%%%%%%%%%%%%
%In this paper, we adopted a (common) choice $\eta=-1$ in eq.~(\ref{6dC}). 
%}
%%%%%%%%%%%%%%%%%%%%%%%%%%%%%%%%%%%%%%%%%


\bibliographystyle{ptephy}
\bibliography{draft_arXiv_v2}

\begin{thebibliography}{10}

\bibitem{Aad:2012tfa}
Georges Aad et~al., Phys. Lett., {\bf B716}, 1--29 (2012),
  {{arXiv:1207.7214}}.

\bibitem{Chatrchyan:2012xdj}
Serguei Chatrchyan et~al., Phys. Lett., {\bf B716}, 30--61 (2012),
  {{arXiv:1207.7235}}.

\bibitem{Libanov:2000uf}
M.~V. Libanov and Sergey~V. Troitsky, Nucl. Phys., {\bf B599}, 319--333 (2001),
   {{arXiv:hep-ph/0011095}}.

\bibitem{Frere:2000dc}
J.~M. Frere, M.~V. Libanov, and Sergey~V. Troitsky, Phys. Lett., {\bf B512},
  169--173 (2001),  {{arXiv:hep-ph/0012306}}.

\bibitem{Neronov:2001qv}
Andrey Neronov, Phys. Rev., {\bf D65}, 044004 (2002),  {{arXiv:gr-qc/0106092}}.

\bibitem{Aguilar:2006sz}
Silvestre Aguilar and Douglas Singleton, Phys. Rev., {\bf D73}, 085007 (2006),
  {{arXiv:hep-th/0602218}}.

\bibitem{Gogberashvili:2007gg}
Merab Gogberashvili, Pavle Midodashvili, and Douglas Singleton, JHEP, {\bf 08},
  033 (2007),  {{arXiv:0706.0676}}.

\bibitem{Guo:2008ia}
Zhi-qiang Guo and Bo-Qiang Ma, JHEP, {\bf 08}, 065 (2008),
  {{arXiv:0808.2136}}.

\bibitem{Kaplan:2011vz}
David~B. Kaplan and Sichun Sun, Phys. Rev. Lett., {\bf 108}, 181807 (2012),
  {{arXiv:1112.0302}}.

\bibitem{ArkaniHamed:1999dc}
Nima Arkani-Hamed and Martin Schmaltz, Phys. Rev., {\bf D61}, 033005 (2000),
  {{arXiv:hep-ph/9903417}}.

\bibitem{Dvali:2000ha}
G.~R. Dvali and Mikhail~A. Shifman, Phys. Lett., {\bf B475}, 295--302 (2000),
  {{arXiv:hep-ph/0001072}}.

\bibitem{Gherghetta:2000qt}
Tony Gherghetta and Alex Pomarol, Nucl. Phys., {\bf B586}, 141--162 (2000),
  {{arXiv:hep-ph/0003129}}.

\bibitem{1126-6708-2000-06-020}
David~Elazzar Kaplan and Timothy~M.P. Tait, Journal of High Energy Physics,
  {\bf 2000}(06), 020 (2000).

\bibitem{Kaplan:2001ga}
David~Elazzar Kaplan and Timothy M.~P. Tait, JHEP, {\bf 11}, 051 (2001),
  {{arXiv:hep-ph/0110126}}.

\bibitem{Huber:2000ie}
Stephan~J. Huber and Qaisar Shafi, Phys. Lett., {\bf B498}, 256--262 (2001),
  {{arXiv:hep-ph/0010195}}.

\bibitem{Kakizaki:2001ue}
Mitsuru Kakizaki and Masahiro Yamaguchi, Int. J. Mod. Phys., {\bf A19},
  1715--1736 (2004),  {{arXiv:hep-ph/0110266}}.

\bibitem{Haba:2006dz}
Naoyuki Haba, Atsushi Watanabe, and Koichi Yoshioka, Phys. Rev. Lett., {\bf
  97}, 041601 (2006),  {{arXiv:hep-ph/0603116}}.

\bibitem{Abe:2008sx}
Hiroyuki Abe, Kang-Sin Choi, Tatsuo Kobayashi, and Hiroshi Ohki, Nucl. Phys.,
  {\bf B814}, 265--292 (2009),  {{arXiv:0812.3534}}.

\bibitem{Csaki:2008qq}
Csaba Csaki, Cedric Delaunay, Christophe Grojean, and Yuval Grossman, JHEP,
  {\bf 10}, 055 (2008),  {{arXiv:0806.0356}}.

\bibitem{Abe:2012fj}
Hiroyuki Abe, Tatsuo Kobayashi, Hiroshi Ohki, Akane Oikawa, and Keigo Sumita,
  Nucl. Phys., {\bf B870}, 30--54 (2013),  {{arXiv:1211.4317}}.

\bibitem{RandjbarDaemi:1982hi}
S.~Randjbar-Daemi, Abdus Salam, and J.~A. Strathdee, Nucl. Phys., {\bf B214},
  491--512 (1983).

\bibitem{Cremades:2004wa}
D.~Cremades, L.~E. Ibanez, and F.~Marchesano, JHEP, {\bf 05}, 079 (2004),
  {{arXiv:hep-th/0404229}}.

\bibitem{Abe:2008fi}
Hiroyuki Abe, Tatsuo Kobayashi, and Hiroshi Ohki, JHEP, {\bf 09}, 043 (2008),
  {{arXiv:0806.4748}}.

\bibitem{Fujimoto:2013xha}
Yukihiro Fujimoto, Tatsuo Kobayashi, Takashi Miura, Kenji Nishiwaki, and Makoto
  Sakamoto, Phys. Rev., {\bf D87}(8), 086001 (2013),  {{arXiv:1302.5768}}.

\bibitem{Abe:2013bca}
Tomo-Hiro Abe, Yukihiro Fujimoto, Tatsuo Kobayashi, Takashi Miura, Kenji
  Nishiwaki, and Makoto Sakamoto, JHEP, {\bf 01}, 065 (2014),
  {{arXiv:1309.4925}}.

\bibitem{Abe:2014noa}
Tomo-hiro Abe, Yukihiro Fujimoto, Tatsuo Kobayashi, Takashi Miura, Kenji
  Nishiwaki, and Makoto Sakamoto, Nucl. Phys., {\bf B890}, 442--480 (2014),
  {{arXiv:1409.5421}}.

\bibitem{Abe:2015yva}
Tomo-hiro Abe, Yukihiro Fujimoto, Tatsuo Kobayashi, Takashi Miura, Kenji
  Nishiwaki, Makoto Sakamoto, and Yoshiyuki Tatsuta, Nucl. Phys., {\bf B894},
  374--406 (2015),  {{arXiv:1501.02787}}.

\bibitem{Matsumoto:2016okl}
Yoshio Matsumoto and Yutaka Sakamura, PTEP, {\bf 2016}(5), 053B06 (2016),
  {{arXiv:1602.01994}}.

\bibitem{Fujimoto:2016zjs}
Yukihiro Fujimoto, Tatsuo Kobayashi, Kenji Nishiwaki, Makoto Sakamoto, and
  Yoshiyuki Tatsuta, Phys. Rev., {\bf D94}, 035031 (2016),
  {{arXiv:1605.00140}}.

\bibitem{Fujimoto:2012wv}
Yukihiro Fujimoto, Tomoaki Nagasawa, Kenji Nishiwaki, and Makoto Sakamoto,
  PTEP, {\bf 2013}, 023B07 (2013),  {{arXiv:1209.5150}}.

\bibitem{Fujimoto:2013ki}
Yukihiro Fujimoto, Kenji Nishiwaki, and Makoto Sakamoto, Phys. Rev., {\bf
  D88}(11), 115007 (2013),  {{arXiv:1301.7253}}.

\bibitem{Fujimoto:2014fka}
Yukihiro Fujimoto, Kenji Nishiwaki, Makoto Sakamoto, and Ryo Takahashi, JHEP,
  {\bf 10}, 191 (2014),  {{arXiv:1405.5872}}.

\bibitem{Cai:2015jla}
Chengfeng Cai and Hong-Hao Zhang, Phys. Rev., {\bf D93}(3), 036003 (2016),
  {{arXiv:1503.08805}}.

\bibitem{Fujimoto:2011kf}
Yukihiro Fujimoto, Tomoaki Nagasawa, Satoshi Ohya, and Makoto Sakamoto, Prog.
  Theor. Phys., {\bf 126}, 841--854 (2011),  {{arXiv:1108.1976}}.

\bibitem{Fujimoto:2016llj}
Yukihiro Fujimoto, Kouhei Hasegawa, Kenji Nishiwaki, Makoto Sakamoto, and
  Kentaro Tatsumi (2016),  {{arXiv:1609.01413}}.

\bibitem{DeWolfe:1999cp}
O.~DeWolfe, D.~Z. Freedman, S.~S. Gubser, and A.~Karch, Phys. Rev., {\bf D62},
  046008 (2000),  {{arXiv:hep-th/9909134}}.

\bibitem{Miemiec:2000eq}
Andre Miemiec, Fortsch. Phys., {\bf 49}, 747--755 (2001),
  {{arXiv:hep-th/0011160}}.

\bibitem{Lim:2005rc}
C.~S. Lim, Tomoaki Nagasawa, Makoto Sakamoto, and Hidenori Sonoda, Phys. Rev.,
  {\bf D72}, 064006 (2005),  {{arXiv:hep-th/0502022}}.

\bibitem{Lim:2007fy}
C.~S. Lim, Tomoaki Nagasawa, Satoshi Ohya, Kazuki Sakamoto, and Makoto
  Sakamoto, Phys. Rev., {\bf D77}, 045020 (2008),  {{arXiv:0710.0170}}.

\bibitem{Lim:2008hi}
C.~S. Lim, Tomoaki Nagasawa, Satoshi Ohya, Kazuki Sakamoto, and Makoto
  Sakamoto, Phys. Rev., {\bf D77}, 065009 (2008),  {{arXiv:0801.0845}}.

\bibitem{Ohya:2010wf}
Satoshi Ohya,
\newblock {SUSY QM meets 5d Gravity},
\newblock In {\em {Supersymmetric quantum mechanics and spectral design.
  Proceedings, Workshop, Benasque, Spain, July 18-30, 2010}} (2010),
  {{arXiv:1012.0301}}.

\bibitem{Nagasawa:2011mu}
Tomoaki Nagasawa, Satoshi Ohya, Kazuki Sakamoto, and Makoto Sakamoto, SIGMA,
  {\bf 7}, 065 (2011),  {{arXiv:1105.4829}}.

\bibitem{Sakamoto:2012ew}
Makoto Sakamoto (2012),  {{arXiv:1201.2448}}.

\bibitem{Williams:2012au}
M.~Williams, C.~P. Burgess, L.~van Nierop, and A.~Salvio, JHEP, {\bf 01}, 102
  (2013),  {{arXiv:1210.3753}}.

\bibitem{Burgess:2012pc}
C.~P. Burgess, L.~van Nierop, S.~Parameswaran, A.~Salvio, and M.~Williams,
  JHEP, {\bf 02}, 120 (2013),  {{arXiv:1210.5405}}.

\bibitem{Witten:1981nf}
Edward Witten, Nucl. Phys., {\bf B188}, 513 (1981).

\bibitem{Cooper:1994eh}
Fred Cooper, Avinash Khare, and Uday Sukhatme, Phys. Rept., {\bf 251}, 267--385
  (1995),  {{arXiv:hep-th/9405029}}.

\bibitem{Cooper:2001zd}
F.~Cooper, A.~Khare, and U.~Sukhatme,
\newblock {\em {Supersymmetry in quantum mechanics}}, 2001).

\bibitem{Cheon:2000tq}
Taksu Cheon, Tamas Fulop, and Izumi Tsutsui, Annals Phys., {\bf 294}, 1--23
  (2001),  {{arXiv:quant-ph/0008123}}.

\bibitem{Nagasawa:2008an}
Tomoaki Nagasawa, Satoshi Ohya, Kazuki Sakamoto, Makoto Sakamoto, and Kosuke
  Sekiya, J. Phys., {\bf A42}, 265203 (2009),  {{arXiv:0812.4659}}.

\bibitem{Lim:1990bp}
C.~S. Lim, Phys. Lett., {\bf B256}, 233--238 (1991).

\bibitem{Lim:2009pj}
C.~S. Lim, Nobuhito Maru, and Kenji Nishiwaki, Phys. Rev., {\bf D81}, 076006
  (2010),  {{arXiv:0910.2314}}.

\bibitem{Kobayashi:2016qag}
Tatsuo Kobayashi, Kenji Nishiwaki, and Yoshiyuki Tatsuta, JHEP, {\bf 04}, 080
  (2017),  {{arXiv:1609.08608}}.

\end{thebibliography}
%
% once the .bbl file has been generated then place the text in your article.

%\begin{thebibliography}{9}

%\bibitem{1}
%J. P. Blaizot, and E. Iancu, Phys. Rep. {\bf 359}, 355 (2002).

%\end{thebibliography}

%\appendix

%\section{Appendix head}

\end{document}